\documentclass[10pt,aps,prf,reprint,onecolumn,nofootinbib,superscriptaddress]{revtex4-2}
\usepackage[linkcolor = blue, citecolor = blue, urlcolor = blue, colorlinks = true]{hyperref}
\usepackage[usenames,dvipsnames]{xcolor}
\usepackage[normalem]{ulem}
\usepackage{graphicx}
\usepackage{listings}
\usepackage{stmaryrd}
\usepackage{amssymb}
\usepackage{amsmath}
\usepackage{gensymb}
\usepackage{float}
\usepackage{bbold}
\usepackage{bm}

\hypersetup{
    colorlinks=true,
    linkcolor=blue,
    citecolor=red}

\begin{document}

\title{Soft biological and bio-inspired materials}

\author{Livio Nicola Carenza}
\affiliation{Department of Physics, Koc University, Rumelifeneri Yolu 34450
Sariyer, Istanbul,  Turkey}
\email{lcarenza@ku.edu.tr}
\author{Giuseppe Negro}
\affiliation{SUPA, School of Physics and Astronomy, University of Edinburgh, Peter Guthrie Tait Road, Edinburgh, EH9 3FD, UK}
\author{Adriano Tiribocchi}
\affiliation{Istituto per le Applicazioni del Calcolo, Consiglio Nazionale delle Ricerche, Via dei Taurini 19, Rome, I-00185, Italy}
\affiliation{INFN Tor Vergata, Via della ricerca scientifica 1, Rome, I-00133, Italy}

\date{\today}

\begin{abstract}
Soft biological materials are a fascinating kind of matter characterized by the unique ability of consuming internal energy to perform complex tasks to preserve the life of the single organism or a whole community. Moreover, biologically inspired materials able to mimic specific features of biological systems can be exploited to design novel materials with vast potential applications in areas ranging from healthcare and robotics to energy and environmental sustainability. 
In this review, we focus on the modeling approaches used to capture the dynamics of this peculiar kind of active matter, starting from the observation of orientational symmetries and reviewing the basis of the so-called active liquid crystal theory, a celebrated model that was successfully used to interpret and investigate a number of different phenomena occurring in both biological and artificial realizations. 
We review the role of topological defects in biological systems and discuss the complex phenomenon of active turbulence, a chaotic dynamical regime observed in a variety of different realizations of systems of both biological and synthetic origin. Finally, we discuss how geometrical and topological confinement can also play an important role in many biological processes, from cellular apoptosis in epithelial tissue, to animal morphogenesis.
\end{abstract}

\maketitle

\section{Introduction}
\label{sec:I}

Soft biological matter refers to a fascinating class of materials that exists at the interface between Biology and Physics. Biological systems are a peculiar kind of soft matter which exhibit unique dynamical and structural characteristics, arising from their intrinsic complexity~\cite{marchetti2013,Bechinger2016,ramaswamy}. 
Contrary to classical equilibrium soft matter which ineluctably evolve to minimize some kind of thermodynamic potential, biological systems evolve out of equilibrium. They are able to convert internal energy -- stored in biomolecules in the form of chemical energy, for instance -- to perform a variety of complex operations which are necessary for the life of a single organism or of a whole colony. 

Indeed, biological matter encompasses a variety of different systems, ranging from the microscopic to the macroscopic scales~\cite{marchetti2013}. For instance, at subcellular scales, biomolecules and proteins together with cytoskeletal filaments cooperate to guarantee the correct cellular functioning and metabolism, but also provide the mechanical structure which allows cells to move and divide~\cite{Needleman2017}.
Single cells can in turn aggregate into multicellular systems. In the case of bacteria, which are single-cell organisms, they can form colonies~\cite{Flemming2016}. Instead, eukaryotes cells often exhibit a more complex degree of organisation, with single cells gathering to form tissues. Different tissues come together to form organs, and different organs to form a whole organism~\cite{Xi2019}. Finally, at macroscopic scale, large assemblies of animals, with many peculiar collective behaviours, fall within the category of biological matter,
as in the case of schools of fish, flocks of birds, or humans in traffic~\cite{cavagna2014,corbetta2023}.

Because of the high degree of biological complexity proper of each particular realisation, Biology and Physics have remained two separate fields for a long time. Nevertheless, starting from the seminal work of Vicsek on flocking~\cite{vicsek1995novel,Vicsek12}, it has become progressively clear that the quantitative and modelling approach of Physics could provide meaningful insights into the behaviours of many biological systems. 
More specifically, the Vicsek model managed to qualitatively capture the dynamical evolution of a flock of birds with a minimal mathematical description, by retaining only a limited set of dynamical variables -- the position and the orientation of each model particle -- therefore minimizing the high biological complexity of the flocking of birds -- a phenomenon whose enchanting dynamics was in ancient times considered as the manifestation of the will of Romans' gods. The celebrated Vicsek model rapidly gained much popularity in the Physics community (among many other models developed to capture collective motion in biological systems), but most importantly paved the way to an entirely novel field of study, that of \emph{Active Matter}~\cite{marchetti2013,Gompper2020}-- a term which is commonly used to define both biological and bio-inspired artificial realizations whose elementary constituents are able to autonomously consume energy, and inherently evolve out of equilibrium.

Fostered by the increasing interest of the community, much effort was spent in the last three decades to decipher the physics beyond many different biological processes, at various levels of organizations, from bacterial biofilm formation, to tissue dynamics and animal collective motion. 
Understanding the mechanical, chemical, and structural properties of soft biological matter has become a prominent area of research, providing an unprecedented development in the field, and unlocking new insights into biological processes, disease mechanisms, and the design of innovative bio-inspired materials and technologies.
Indeed, drawing inspiration from the remarkable adaptability and functionalities of living organisms, advanced synthetic materials may be developed to mimic properties of biological systems, including self-assembly, shape-shifting, and self-regulation. Biologically inspired soft matter holds vast potential in diverse applications, ranging from healthcare and robotics to energy and environmental sustainability~\cite{Ghosh2020,Fu2022}.

This review explores the captivating world of soft biological matter, and it is organised as follows.
In Section~\ref{sec:II} we provide an overview of some iconic biological and bio-inspired active systems, with particular attention towards microscopic fluidic realisations.
In Section~\ref{sec:III}, we focus on the occurrence of orientational order in active systems and review the basis of the active liquid crystal theory, a celebrated model successfully used to interpret and investigate a number of different phenomena occurring in both biological and artificial realisations.
In Section~\ref{sec:IV} we examine the relevance of topological defects in the evolution of active systems and how these are exploited by living systems.
In Section~\ref{sec:V} we discuss the importance of geometry and topological confinement as a tool to control the dynamics of active systems. Finally, Section~\ref{sec:VI} will be devoted to an overview of the open problems and future perspective in the field of Active Matter.

\section{Active Matter: from biological aggregates to artificial assemblies}
\label{sec:II}

Active systems, characterized by their ability of consuming internal energy to do work on the surrounding environment, can have both biological and artificial origin. Examples of biological active systems are agglomerates of living organisms, such as bacterial colonies~\cite{Wensink2012,Yaman2019,Copenhagen2021} and cellular tissues~\cite{Saw2017,Kiehart2017,armengol2023}, where individual constituents often cooperate with each other to preserve the life of the whole system by accomplishing to specific tasks. 
Artificial realizations, instead, can be in turn classified between \emph{in vitro} systems -- i.e. systems whose elementary constituents have biological origin and have been prepared and reorganized to perform experiments outside of a living organism -- and synthetic systems, whose constituents have been opportunely engineered to mimic a specific biological behavior. Examples of the former class of systems are suspensions of cytoskletalal filaments, such as microtubules and actomyosin, extracted from cells and prepared in the lab through the addition of chemicals to obtain novel materials with properties unobserved in classical equilibrium soft systems~\cite{sanchez2012,Keber2014,Guillamat2018,kumar2018,Martinez2021}, while examples of the latter are active colloids and Janus particles, which are designed to convert chemical into kinetic energy~\cite{Kiehart2017,Bechinger2016}.
In this Section, we will provide an overview of the aforementioned systems, focusing on their peculiar dynamical regimes.

\begin{figure}[t!]
\centering
\includegraphics[width=1.0\columnwidth]{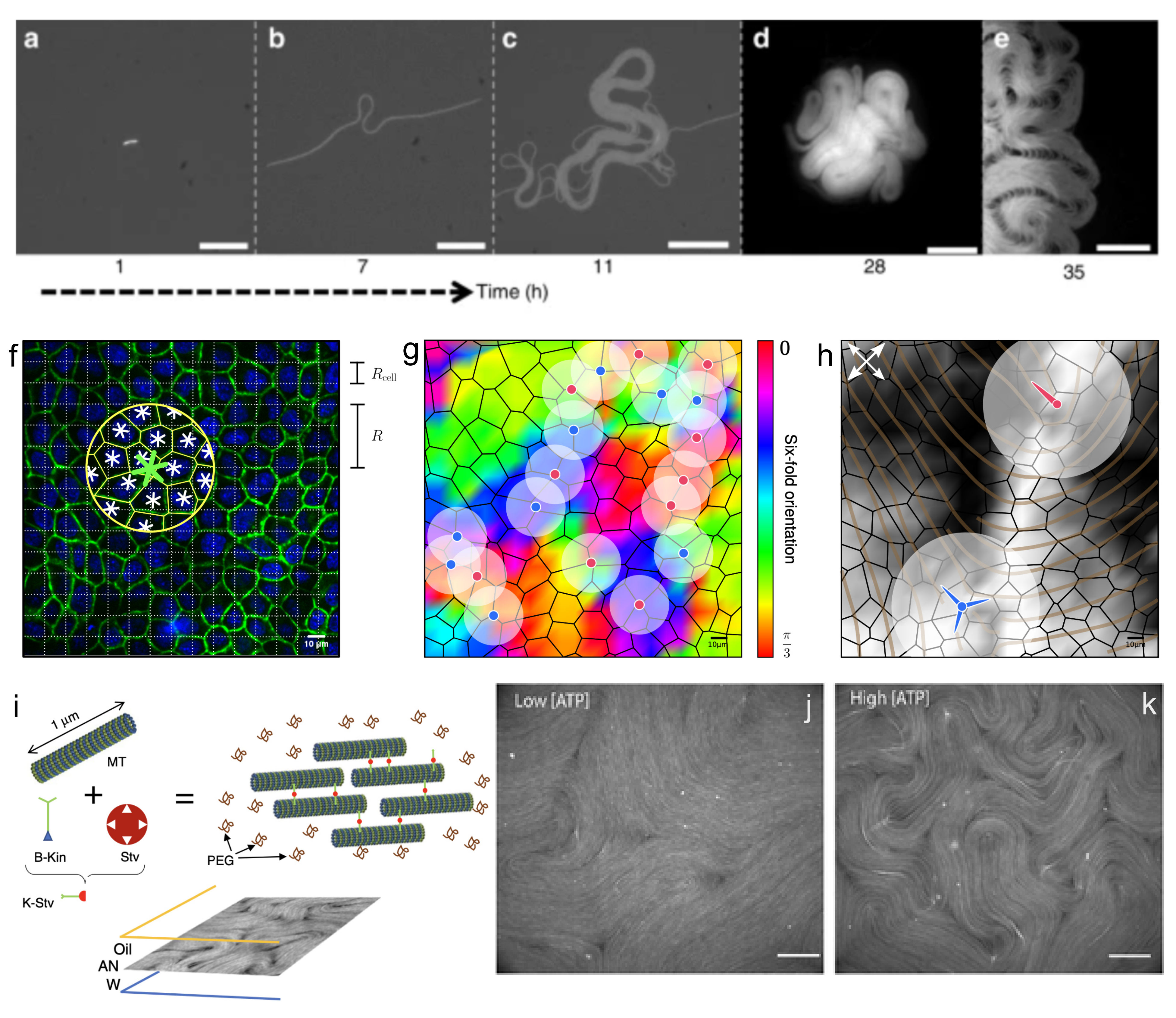}
 \caption{\textbf{Bacterial colonies, epithelial tissues, active gels.} (a-e) Growth of a biofilm of \emph{B. subtilis} starting from a single bacterium. As the bacteria proliferate, they form long chains which develop buckling instability (b-c) until a multilayered structure is fully formed (d-e), characterized by nematic order and nematic topological defects. Scale bars $30 \mu{\rm m}$. Adapted from Ref.~\cite{Yaman2019}. (f-h) Experiments of Madin-Darby canine kidney (MDCK) epithelial tissues feature multi-scale hexanematic order. Hexatic and nematic cell-wise orientation (white stars in panel (f)) can be coarse-grained to define the local direction of alignment (green star). Hexatic symmetry is dominant at small coarse-graining scales (panel (g)) while nematic symmetry prevails at larger scales (panel (h)). Adapted from Ref.~\cite{armengol2023}. (i-k) Microtubule suspensions at a 2D water-oil interface in the presence of a depletion agent, such as poly-ethylene glycol (PEG), develop nematic textures. The suspension can be \emph{activated} with ATP, whose concentration variation controls the dynamical response of the active fluid and its morphological features. 
 Panel (j) and (k) show two images at low and high concentration, respectively. Scale bars $100 \mu m$. Panel (i) adapted from \cite{doostmohammadi2018}; panels (j) and (k) adapted from~\cite{Lemma2019}.}
 \label{fig:1}
\end{figure}

\paragraph*{\bf Bacterial colonies.}
Bacterial suspensions have emerged as invaluable experimental systems in the study of active matter, captivating the attention of the physics community due to their inherent self-propelled nature and (relatively) simple interactions with the environment and other entities~\cite{Wensink2012}. Representing a natural realization of self-propelled particles, bacteria offer an elementary yet relevant example of how active constituents can lead to self-assembly and the emergence of complex patterns~\cite{Yaman2019}. These microorganisms exhibit enhanced chemotactic properties, enabling them to respond to external stimuli, such as temperature variations, nutrient availability, and oxygen concentration, giving rise to diverse colonies ranging from biofilms to fluidic suspensions~\cite{Wadhams2004,Dombrowski2004}. Remarkably, the uncoordinated evolution of individual bacteria alone can drive the development of elaborate patterns in these systems. By controlling environmental factors such as temperature, oxygen availability, or light intensity, the typical speed of bacterial migration can be tuned, leading to intriguing dynamical regimes. For instance, dense suspensions of flagellate bacteria in fluids can arrange into liquid crystalline patterns and enter a chaotic flowing state known as bacterial turbulence, characterized by whirling patterns reminiscent of fluid turbulence at high Reynolds numbers~\cite{Sokolov2007,Zhang2010,Wensink2012,Dunkel2013,Copenhagen2021}. 
When faced with adverse factors such as limited nutrients, temperature fluctuations, or the presence of toxic substances (such as antibiotics), individual bacteria join forces and adhere to surfaces, forming a protective community known as a biofilm~\cite{Wadhams2004}.
Biofilm formation is indeed a remarkable survival strategy employed by bacteria to overcome challenging environmental conditions through primordial social interactions. 
Within the biofilm, bacteria engage in intricate social behaviours, exchanging chemical signals, and communicating through a process called quorum sensing~\cite{Miller2001}. This communication enables the development of specialised functions such as the production of extracellular matrix that provides structural integrity and protection, accompanied by the development of ordering features, as shown in Fig.~\ref{fig:1}(a-e) for the case of a biofilm of \emph{B. subtilis} exhibiting the typical characteristics and dynamical instabilities of a nematically ordered system. The fascinating process of biofilm formation demonstrates the power of collective behavior and highlights the adaptability and resilience of bacterial communities, paving the way for further exploration of social interactions in microbial ecosystems and their implications in medicine, environmental science, and biotechnology.

\paragraph*{\bf Epithelial Tissues.}
Epithelial tissues are fundamental in animal organisms, playing crucial roles in maintaining organs integrity, serving as barriers, and facilitating interactions with the environment~\cite{Friedl2009}. They are are made up of tightly packed cells that form continuous layers~\cite{depascalis2017} that line the surfaces of organs, cavities, and blood vessels throughout the body. The primary function of epithelial tissues varies depending on their location; for example, they provide protection to underlying tissues in the skin, absorb nutrients in the intestine, and secrete substances in glands. 
Their ubiquity in animal organisms makes them central in a number of biological processes which are fundamental for the correct functioning of animal life. 

Of particular relevance is embryogenesis, i.e. the process of animal development which, starting from a single fertilised egg, leads to the development of an independent organism. During embryogenesis, epithelial tissues undergo dynamic changes, through a series of cell shape modifications, migrations, and rearrangements, ultimately shaping the overall architecture of the developing organism.
Indeed, embryogenesis is a remarkable example of how epithelial tissues can orchestrate complex morphogenetic processes. An intriguing model system, widely studied in the literature,  is the fruit fly (Drosophila melanogaster) embryo which, after the first initial stage of development where cells divide at a high rate, undergoes a series of precisely coordinated cell divisions and rearrangements, leading to the formation of distinct epithelial sheets, the so-called germ layers, namely the ectoderm, mesoderm, and endoderm. These germ layers give rise to different tissues and organs in the developing embryo, setting the foundation for the complex body plan of the organism~\cite{Streichan2018}.
This process, known as \emph{gastrulation}, consists of a sequence of well-coordinated cellular movements, such as convergence, extension, and invagination~\cite{Martin2020}. 
Studying the dynamics of epithelial tissues in the fruit fly embryo provides valuable insights into the principles of tissue mechanics and cellular interactions that govern embryonic development. 
A further process where the dynamics of epithelial tissues plays a central role in tissue repair and regeneration is the wound healing. Here, epithelial cells surrounding a wound site undergo rapid migration and proliferation to close the gap and restore tissue confluency, with a coordinated movement  (known as cell crawling) driven by the contractility of their cytoskeleton and guided by signals from the surrounding microenvironment, which enable cells to \emph{sense} the presence of a boundary (i.e the wound)~\cite{SerraPicamal2012}. Understanding the biophysical and molecular mechanisms underlying epithelial cell crawling and tissue remodeling during wound healing is essential for developing novel therapeutic approaches to enhance tissue repair and accelerate healing.

Apart from being central to many beneficial biological processes, epithelial tissues are also deeply involved in life-threatening phenomena, as in the case of a metastatic cascade -- the process by which cancer cells spread from the primary tumour to distant organs. The metastatic cascade involves a series of complex steps, including the invasion of cancer cells through surrounding tissue, the entry into the bloodstream or lymphatic vessels, and extravasation into distant tissues~\cite{Haeger2019}. Interestingly, epithelial-to-mesenchymal transition is a key process that occurs during metastasis, whereby epithelial cells lose their cell-cell adhesions and acquire a migratory and invasive mesenchymal phenotype~\cite{Sciacovelli2017}. This transition allows cancer cells to leave the primary tumour and invade surrounding tissues, facilitating their dissemination to distant sites. Shedding light on the mechanisms underlying epithelial tissue plasticity and how they relate to cancer metastasis may be crucial for designing targeted strategies to impede the spread of cancer cells.
Finally, it is important to mention that in {\it vitro} cell monolayer experiments (Fig.~\ref{fig:1}(f-h)) are highly implemented in epithelial tissue research because they provide a simplified yet controlled model system for investigating fundamental cellular and tissue behaviours. Indeed, by culturing epithelial cells in a single layer on a flat surface, it is possible to study cell-cell interactions, migration, proliferation, and responses to external stimuli, offering insights into the dynamic behaviors of epithelial tissues in health and disease~\cite{Saw2017,Arslan2021,Balasubramaniam2021,armengol2023}.

\paragraph*{\bf Active Gels: Microtubules and Actomyosin suspensions.}
Aside from living biological systems, latest research has focused on systems derived from biological constituents, but artificially assembled to obtain new materials with striking structural and dynamical properties~\cite{doostmohammadi2018}. This is the case of microtubules~\cite{Keber2014,Guillamat2018} and actomyosin suspensions~\cite{kumar2014,kumar2018}, which have become valuable model systems in the study of active matter, providing insights into the collective behaviors and dynamics of self-propelled entities. These suspensions are typically obtained by confining microtubules and actomyosin filaments, extracted from cells, at a two-dimensional interface between oil and water, for instance, together with some depletion agent (a non-absorbing polymer like PEG) to drive the formation of filament bundles~\cite{sanchez2012} (see Fig.~\ref{fig:1}(i)). To \emph{activate} these systems, ATP (adenosine triphosphate, an organic compound used in living cells as a source of energy) is added, providing the necessary energy for motor proteins to bind and move along the filaments. As a result, the filaments exhibit self-propulsion, leading to a rich array of dynamical phenomena.

A striking feature of such systems is the appearance of active liquid crystal phases~\cite{julicher2007,sanchez2012,prost2015,doostmohammadi2018,Pismen2021}, which are characterised by aligned filaments akin to liquid crystals but with the added complexity of self-propulsion. In this state, the filaments display a common orientation, leading to coherent motion over large length scales. The competition between alignment and active motion can give rise to topological defects, such as disclinations (Fig.~\ref{fig:1}(j-k)), which are a fingerprint of the existence of liquid crystalline order, as we will see in more detail in the next Section. 
Another mesmerising property of active suspensions is the occurrence of \emph{active turbulence} -- a particularly intriguing dynamical regime observed at high activity, when filament densities and ATP concentrations are sufficiently large~\cite{Alert2022}. Under these conditions, the energy injection provided by ATP hydrolysis leads to the proliferation of topological defects, so that the resulting dynamics develops complex chaotic features, displaying vortices, whirls, and disordered flows reminiscent of classical fluid turbulence. This chaotic state arises from the complex interplay between the self-propulsion of filaments and their elasticity, leading to large-scale collective motion and energy dissipation through friction and viscosity. A more detailed discussion on active turbulence is given in Section~\ref{sec:IV-B}.

Confined geometries introduce additional constraints leading to unique behaviors. 
In microchannels, for example, confinement of active gels can lead to novel dynamical states, such as spontaneous flows and dancing disclinations \cite{voituriez2005,Shendruck2017,Hardouin2022};
when confined to fluid droplets, active gels can yield autonomous motility \cite{mogilner1996,theriot2008,poincloux2011,aronson2012,sanchez2012,tjhung,tjhung2015,tiribocchi2023,tiribocchi2023pof} and spontaneous division \cite{giomi2014} while, on thin spherical shells, they display periodic self-synchronised behaviours or develop spontaneous curvature-induced morphologies~\cite{Keber2014,Zhang2016,Guillamat2018}

In general, microtubule and actomyosin suspensions in active matter experiments provide a versatile and controllable platform to study emergent collective behaviours. By manipulating ATP concentrations, filament densities, and confining geometries, researchers can explore a wide range of dynamical regimes and gain fundamental insights into the principles governing active matter systems. In addition, these systems 
shed light on the dynamics of biological processes involving cytoskeletal elements and inspire the design of novel synthetic active materials and technologies, with applications ranging from soft robotics to advanced materials engineering \cite{Needleman2017}.

\paragraph*{\bf Colloids and Janus particles.}
\begin{figure}[t!]
\centering
\includegraphics[width=1.0\columnwidth]{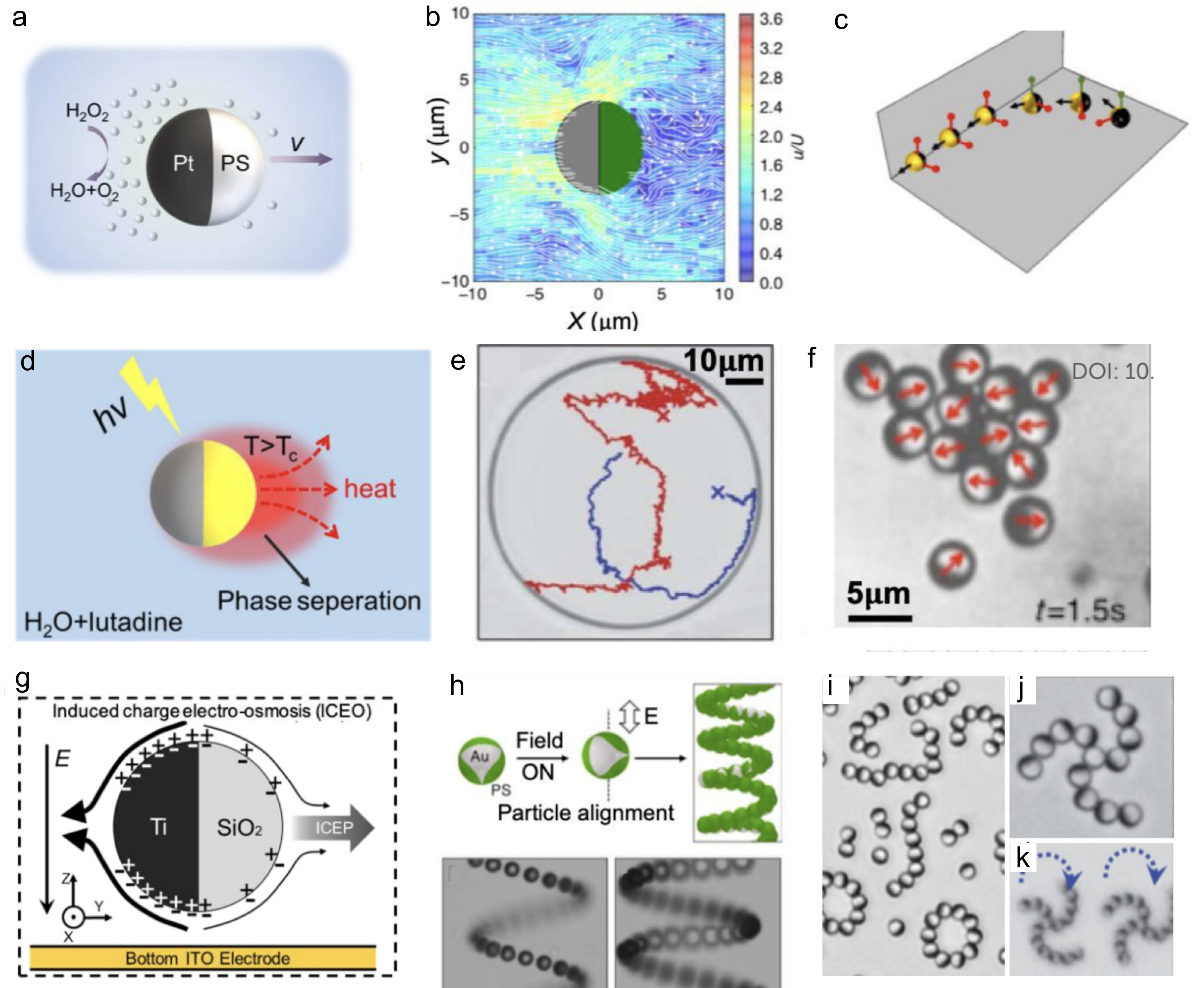}
 \caption{\textbf{Active Colloids.} (a) ${\rm Pt}$-coated spherical microswimmers moving by asymmetrically catalysing the decomposition of ${\rm H}_2{\rm O}_2$ into water and oxygen. Adapted from  Ref. \cite{reviewcolloidsmechanisms}.  (b) Experimentally measured flow field around a ${\rm Pt}$-coated swimmer in ${\rm H}_2{\rm O}_2$. Adapted from Ref.~\cite{Campbell2019}.
 (c) A ${\rm Pt}$-coated swimmer preferentially moves close and along a wall. Adapted from Ref.~\cite{Das2015}.  (d) Janus microswimmers demixing water-lutidine.  These microswimmers heat up asymmetrically under light, and locally demix a critical binary mixture of water and lutidine. Adapted from Ref. \cite{reviewcolloidsmechanisms} (e) Microswimmers preferentially move along walls of a circular confinement. Reprinted from~\cite{Volpe2011} . (f) Microswimmers spontaneously form clusters. Reprinted from Ref.~\cite{Buttiononi2013}.
 (g) Operating mechanism of ICEP, arising from asymmetric polarization of the two hemispheres of a metal-dielectric Janus microsphere under an electric field. Adapted from Ref. \cite{reviewcolloidsmechanisms} (see main text for details). (h) A microsphere with gold patches experience torques due to ICEP under an AC electric field, and moves in controlled, helical trajectories. Reprinted from \cite{Lee2019}.  (i-k)  Dynamic assembly of ICEP microswimmers into interacting chains.  Reprinted with permission, (i) from \cite{Nishiguchi_2018} and (j)-(k) from \cite{zhangfaraday_2016}.
 }
 \label{fig:2}
\end{figure}
Having reviewed living and biological realizations of active matter, we now focus on  synthetic systems that utilize external energy inputs to influence the structure and behavior of material components~\cite{Kiehart2017,Bechinger2016,Bishop2023}. 
In this context, colloidal dispersion of solid particles suspended in liquid mediums are particularly useful building blocks, covering a size range of $0.1 - 10 \mu {\rm m}$.
These systems have a rich history as experimental models for condensed matter, 
serving as analogs to \emph{large atoms}~\cite{poon2004} that, interacting and coming together, create various condensed phases, including crystalline solids, disordered liquids, glasses, and gels~\cite{Lu2013,Joshi2014,Zaccarelli2009}. 
These passive particles can be activated through catalytic reactions or external fields, turning them into colloids capable of autonomous movement and spontaneous assembly into dynamic structures, unhindered by the constraints of thermodynamic equilibrium~\cite{wang2013,Bechinger2016,zhang2017}. Here, we review some common active colloid systems and three main mechanisms of self-propulsion.

\emph{Janus particles.} Arguably the most studied active colloids is the type powered by surface chemical reactions. These particles are often spheres or rods with a Janus structure -- named after the two-faced Roman god Janus -- and  generally consist of two opposite-facing segments  undergoing either electrophoresis or diffusiophoresis, as a result of an asymmetric chemical reaction ~\cite{Palacci2014}. It was the introduction of Platinum (${\rm Pt}$) coated Janus microspheres moving in ${\rm H}_2{\rm O}_2$ that truly kindled interest among physicists~\cite{Campbell2019}.
Since then, this type of active colloid has quickly emerged as a popular choice for model systems, thanks to its ease of fabrication. The self-propulsion mechanism in this case originates from the catalytic decomposition of ${\rm H}_2{\rm O}_2$ into water and oxygen on the surface of ${\rm Pt}$, giving rise to a electrophoretic mechanism that ultimately propels the colloids (Fig.~\ref{fig:2}a-b)~\cite{Campbell2019}. These systems, for example, show an interesting dynamics in complex environments with obstacles, such as  a preferential propulsion along walls or trenches (Fig.~\ref{fig:2}c)~\cite{reviewcolloidsmechanisms,Das2015,Simmchen2016}.

\emph{Light-activated colloids.} Using light as an activation strategy confers to this class of colloids unique benefits, such as easy external power and control and excellent temporal modulation~\cite{Zhang2018ul}.  
In this case, photothermal effect or photochemical reaction are  exploited in the design~\cite{Palacci2014}. Their propulsion mechanisms essentially consists of 
three processes (Fig. 2d). First, the metal (or carbon) cap of a Janus particle absorbs light and releases heat, which increases the local temperature around the cap. Then, Janus particles are suspended in a critical binary mixture of water and $2,6$-lutidine of a critical temperature of $307 K$, above which the liquid demixes into water and lutidine phases. In the third and final step, the demixed water and lutidine accumulates preferentially at the hydrophilic and hydrophobic surfaces, respectively. This concentration gradient and the difference in the interactions with the surface move the particle via self- diffusiophoresis~\cite{Araki2019}.
These types of light-powered Janus colloids have been used as a model system to study the dynamical and morphological properties in complex and crowded environments~\cite{Bechinger2016,PATTESON201686,Li2016}, such as among microstructures where they show preferential attachment to walls    (Fig.~\ref{fig:2}e)~\cite{Volpe2011}, as well as  
the cluster formation and collective behaviours in systems of active colloids. In this respect, different studies have shown that these colloids exhibit motility induced phase separation (MIPS)~\cite{CatesReview2015,gonnella2014phase,lei2019nonequilibrium, Gonnella15,Caporusso22,Caporusso20,Solon15,KlKaKr18,Negro2022EPJE,Caporusso2024SoftMatter,Klamser2021,Cates_2013}, a process where colloids phase separate into large clusters in a dilute gas phase at large number densities and in the absence of attractive interactions (Fig.~\ref{fig:2}f)~\cite{Buttiononi2013}.

\emph{Electrically powered self-propelled particles.} Electrically powered active colloids are characteristic realizations of self-propelled particles that hold very promising application in the design of novel materials~\cite{Bazant2004,BAZANT2010203,SQUIRES_BAZANT_2004}. The main mechanism of self propulsion is, in this case, the induced-charge electrophoresis (ICEP) (Fig.~\ref{fig:2}g)~\cite{SQUIRES_BAZANT_2006,Yariv2005,Boymelgreen2015,Boymelgreen2014,Shields2018}, which exploits electrons redistribution in an electrically conducting microparticle in the presence of an external electric field, inducing an electric dipole. This dipole leads to a non-uniform surface charge which attracts, from the solution, charged ions having the opposite polarity to the local surface charge. This, in turn, couples to the external electric field and produce, on the particle surface, an electro-osmotic flow much stronger at the metal side  (because the metal cap polarizes more intensely than the dielectric cap). The Janus particle therefore moves by the imbalanced flow, with the dielectric end leading. Devising  Janus particles with metal patches of particular shapes and asymmetry can lead to spontaneous and chiral rotation, providing a  strategy to steer a swimmer in arbitrary trajectories by switching its dynamics between a straight \emph{run} and a rotating \emph{tumble} phase~\cite{Mano2017}, or control helical trajectories in 3D under AC electric fields (Fig.~\ref{fig:2}h)~\cite{Lee2019}. 
Another notable feature of ICEP swimmers is their pairwise interactions, such as  dipolar interactions. They arise from the electrical polarization of the colloidal particles, and are repulsive for those polarized in the same directions and attractive otherwise. Importantly, since this polarization depends on materials and AC frequency, the dipolar interaction between two colloids can be switched between attraction and repulsion by varying driving frequencies, a process leading  
to collective states such as swarms, chains, clusters and isotropic gases~\cite{Yan2016}, and flagella-like chains of Janus particles (Fig.~\ref{fig:2}h-k)~\cite{Nishiguchi_2018,Lee2019}. 
Finally, at high density, a strong pairwise repulsion between ICEP swimmers can give rise to active turbulence~\cite{Nishiguchi2015}.

Beyond their role as model system, active colloids hold a great potential as building blocks in  a wealth of applications, from microrobotics~\cite{Kohee2018,Palagi2018} to smart materials~\cite{Salazar2021}. As mobile robots, these self-propelled particles hold the promise of navigating intricate micro-environments to manipulate both matter and information in precise locations that would otherwise be inaccessible.
For instance,
microrobots engineered for autonomous cargo retrieval should possess the capability to respond to local flow gradients, capture specific cargo, and return autonomously without external intervention or
control. While living cells routinely perform these fundamental tasks, synthetic robots could effectively function even in harsh environments hostile to life,
a capability that could revolutionize precision diagnostics, therapeutic delivery, and noninvasive surgery in the realm of human health.
They also have potential applications in environmental cleanup and material restoration, where colloidal robots can seek out and neutralize toxic substances or repair structural flaws. Achieving these ambitious objectives will necessitate significant advancements in our ability to design active colloids with intelligent behaviors that adapt to their immediate surroundings \cite{tevrugt}. 

\section{Orientational Order and Active Liqid Crystals}
\label{sec:III}

In the previous Section we have provided an overview of some typical active soft materials which significantly differ for their origin and preparation. 
Nevertheless, many active systems share an important feature which is at the base of many models developed to describe active matter, i.e. the emergence of orientational order, where the orientations of individual entities become correlated over length-scales significantly larger than the typical size of the individual constituents.
This feature earned active
systems with orientational order  the name of active liquid crystals.

To elucidate this concept, we start from the case of a two-dimensional bacterial suspension, composed by $N$ model particles, whose position is $\bm{r}_{i}$, being $i=1,\dots,N$ and index labelling each organism. Moreover, we denote with the unit vector $\bm{\nu}_{i} = \cos \theta_{i} \bm{e}_{x} + \sin \theta_{i} \bm{e}_{y}$ their individual orientation. At low concentrations, bacteria display disordered (or isotropic) states where individual entities move in uncorrelated directions with respect to each other, resulting in no orientational order. However, as the density of the constituents increases, a transition to an ordered phase occurs, where the individual self-propelled entities tend to mutually align their orientation due to interactions mediated by their motility. Such process is akin to a phase transition and a hallmark of many active matter systems.
Importantly, this means that, instead of describing the dynamics of the single particles, we can coarse-grain bacterial orientations, by averaging the orientation of those particle whose position falls within the coarse-graining length-scale $R_{\rm cg}$ and define a mesoscopic \emph{polarization} field at position $\bm{r}$ as
\begin{equation}
\bm{p}(\bm{r}) = \dfrac{1}{N_{\rm cg}} \sum_{i=1}^{N} \bm{\nu}_i \vartheta_{H}(R_{\rm cg}-|\bm{r}_i-\bm{r}|). 
\label{eq:1}
\end{equation}
Here, $N_{\rm cg} = \sum_{i} \vartheta_{H}(R_{\rm cg}-|\bm{r}_i-\bm{r}|)$ denotes the number of particles which fall within the coarse-graining disk, with $\vartheta_{H}$ the Heavside step function, such that $\vartheta_{H}(x)=1$ if $x \geq 0$ and $0$ otherwise. 
We stress that the coarse-grained order parameter $\bm{p}(\bm{r})$ captures both the average direction of alignment through its direction and the magnitude of local orientational order through its modulus, which is $0$ for an isotropic system and $1$ for a perfectly aligned state. 

Interestingly enough, there are situations where the polar order parameter becomes null (or significantly small) and yet it is possible to define a notion of orientational order. In bacterial systems this happens, for instance, during the process of biofilm development of some bacterial strains, when bacteria mutually align their body irrespective of the head/tail orientation of their neighbours (see Fig.~\ref{fig:1}a-e), so that the right-hand side of Eq.~\eqref{eq:1} identically vanishes, signaling the absence of polar symmetry. 
In this case, we speak of \emph{nematic} symmetry and the emerging orientational order can be captured through the nematic tensor
\begin{equation}
\bm{Q}(\bm{r}) = \dfrac{1}{N_{\rm cg}} \sum_{i=1}^{N} \left( \bm{\nu}_i \bm{\nu}_i - \dfrac{1}{2}\bm{I} \right) \vartheta_{H}(R_{\rm cg}-|\bm{r}_i-\bm{r}|),
\label{eq:2}
\end{equation}
where $\bm{I}$ denotes the identity tensor in $2D$. One can immediately verify that the symmetric and traceless $Q$-tensor is invariant under the inversion of the particle orientation $\bm{\nu}_i \rightarrow -\bm{\nu}_{i}$, and therefore it does not vanish even when $\sum_i \bm{\nu}_i = 0$. We stress that in $2D$, a traceless nematic tensor only has two independent components, therefore it encapsulate the same amount of information as the polarization field of Eq.~\eqref{eq:1}. Indeed, a traceless symmetric tensor can always be represented in $2D$ as $\bm{Q} = S(\bm{n}\bm{n} - \bm{I}/2)$, where $\bm{n}$ is the \emph{director} -- a head-tail symmetric unit vector which defines the local direction of alignment (see Fig.~\ref{fig:3}a,b,c)-- while $S$ measures the degree of orientational order.
The $Q$-tensor is a standard mathematical tool used in the description of nematic liquid crystals \cite{degennes,lavren_book}, both in two- and three-dimensional geometries and it has found numerous applications in active matter.  
Indeed, nematic order is very common in biological matter, and it is observed in bacterial systems, microtubule and actomyosin suspensions, and epithelial tissues \cite{marchetti2013,ramaswamy}.

However, polar and nematic order are not the only example of symmetries observed in biological and bio-inspired systems.
A particularly relevant case observed, for example, in colloidal particles and epithelial tissues (see Fig.\ref{fig:1}f),
is the hexatic order, where the system is invariant under rotation of $2\pi/6$.
For instance, a hexatic orientation can be assigned to cells in tissues by computing the complex shape function $\gamma_6$ from the position of the cell's vertices as  
\begin{equation}
\gamma_{6} = \dfrac{1}{\Delta_{6}} \sum_{v} |\bm{r}_{v}|^6 e^{i6\theta_{v}},
\label{eq:3}
\end{equation}
where $\Delta_{6} = \sum_{v} |\bm{r}_{v}|^6$ is a normalisation factor, with $\bm{r}_{v}$ and $\theta_{v}$ position and orientation of the vertices of the cell with respect to the cell's centre of mass, and $v$ an index labelling each vertex \cite{armengol2023}. 
The phase ${\rm Arg}(\gamma_{6})/6$ of the shape function defines the $6-$fold orientation of whole cell and can be graphically represented as $6-$legged star located at the centre of the cell and oriented in a way that maximizes the probability of finding a vertex in the directions of the legs (see Fig.~\ref{fig:3}f)~\footnote{The shape function $\gamma_6$ can be seen as a generalization of the bond-orientational order parameter commonly used in the context of two-dimensional melting. We suggest the interested reader to refer to Ref.~\cite{armengol2023} for further details.}. Analogously to the polar and nematic order, the cell-wise shape function can be coarse-grained to provide a mesoscopic description of a cellular tissue by defining the shape order parameter
\begin{equation}
\Gamma_{6}(|\bm{r}|) = \dfrac{1}{N_{\rm cg}} \sum_{i} \gamma_{6}(\bm{r}_i) \vartheta_{H}(R_{\rm cg}-|\bm{r}_i-\bm{r}|),
\label{eq:4}
\end{equation}
where the summation is performed over all cells in the system located at position $\bm{r}_{i}$, while $N_{\rm cg}$ represents, once again, the number of cells whose centre of mass falls within the coarse-graining disk with radius $R_{\rm cg}$ at position $\bm{r}$. We also notice that the approach here presented for the shape function $\gamma_{6}$ allows for an elegant generalization to the case of generic $p-$atic symmetry, so that a complex function $\gamma_{1}$ can be defined to capture polar order, $\gamma_{2}$ for nematic order, and so on \cite{armengol2023}, with a  
univocal correspondence between these and the polar and nematic order parameter introduced earlier in this Section.
As a final remark, we observe that some biological systems can exhibit more than just one single orientational symmetry. This is the case, for example, of MDCK epithelial cells which, when 
cultured in {\it vitro} on uncoated glass, exhibit both nematic and hexatic order, with the former dominant at large length-scales and the latter at small length-scales \cite{armengol2023}.

\subsection{Topological defects}
\label{sec:III-A}
\begin{figure}
\centering
\includegraphics[width=1.0\columnwidth]{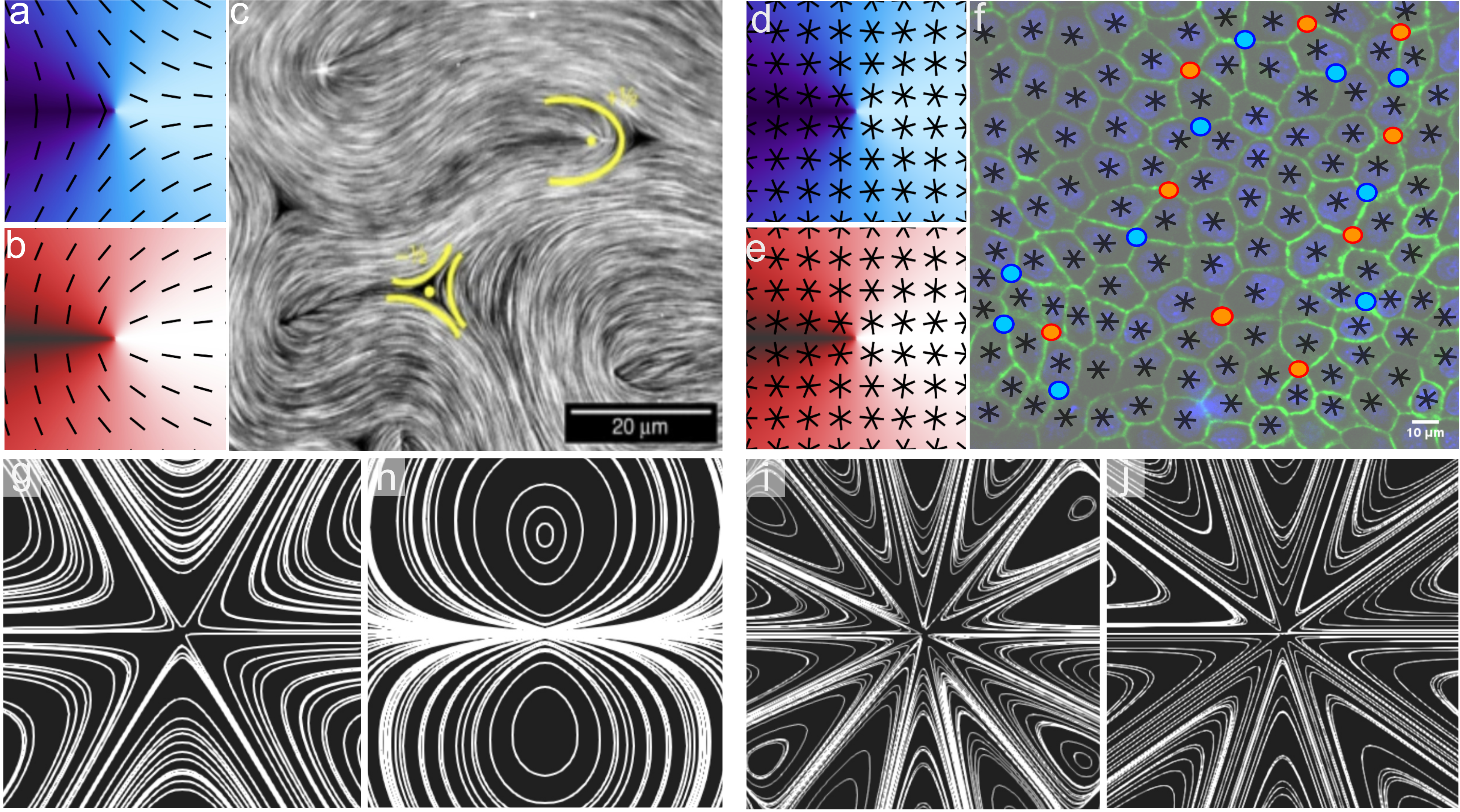}
 \caption{\textbf{Topological Defects.} (a-b) Cartoon of a negative $-1/2$ (panel~a) and positive $-1/2$ (panel~b) disclination with associated nematic director field. (c) Disclinations (highlighted in yellow) in an active nematics in a 2D microtubule suspension. (d-e) Cartoon of a negative $-1/6$ (panel~d) and positive $+1/6$ (panel~e) hexatic disclination. (f) Positive (red circles) and negative (blue circles) hexatic defects in experiments of MDCK cell monolayers. Six-legged stars show the cell-wise reconstructed 6-fold orientation according to Eq.~\eqref{eq:3}. (g-j) Stream-lines of the passive backflow generated by elastic stresses for the negative (panel~g) and positive (panel~h) nematic defects shown in panels a-b, and for the negative (panel~h) and positive (panel~i) hexatic defects shown in panels d-e. Panels a,b,d,e,g-j adapted from Ref.~\cite{Krommydas2023}; panel~c adapted from Ref.~\cite{doostmohammadi2018}; panel~f adapted from Ref.~\cite{armengol2023}.}
 \label{fig:3}
\end{figure}
In the previous Section we have reviewed the possible orientational symmetries featured by biological and bio-inspired materials and discussed how the mathematical description typically used to characterise order in classical liquid crystal systems can be exploited to capture order in active systems. 
Active liquid crystals often feature properties analogous to their passive counterpart, as for instance elastical properties, although, irrespective of the $p-$atic order of orientational symmetry, the locally ordered pattern of active units can deform giving rise to continuous deformations. In passive liquid crystals, these are associated to an energy cost which, in the framework of the Landau-de Gennes theory \cite{degennes}, can be quantified in terms of the free-energy
\begin{equation}
\mathcal{F}^{\rm LdG} = \int {\rm d}\bm{r} \, \left[ - \dfrac{A_{\rm bulk}}{2} |\Gamma_{p}|^{2} + \dfrac{A_{\rm bulk}}{4} |\Gamma_{p}|^{4}  + \dfrac{L_{p}}{2}(\nabla \Gamma_{p})^2 \right],
\label{eq:5}
\end{equation}
where we have used the notation for a liquid crystal of generic $p-$atic order. Here, $A_{\rm bulk}$ is a material parameter that drives the transition to the order state with $|\Gamma_{p}|=1$ and $L_{p}$ is the elastic constant of the (active) liquid crystal. 

While in passive systems distortions of the liquid crystal are minimised to reduce the elastic cost and achieve a globally aligned ground state, in active systems the internal energy injection has the property of counteracting such a relaxation process and enhancing the deformations. This effect can dramatically affect the liquid-crystalline arrangement and give rise to singularities -- the \emph{topological defects} -- where it is no longer possible to define a notion of orientational order~\footnote{We stress that topological defects are an ubiquitous feature of many classical systems. According to the particular system under consideration, they are classified as vortices, dislocations, disclinations, \emph{etc.} Furthermore, topological defects can also be found in steady state in classical systems as the result of thermal fluctuations or specific boundary conditions \cite{MerminRevModPhys}.}. 
Therefore, a topological defects in $2D$, also known as \emph{disclination}, represents a point where the local orientation of alignment cannot be determined unambiguously and is characterised by a core region where the molecular alignment changes gradually around a central point. 
It is worth mentioning that while in $2D$ topological defects are pointwise, in $3D$ they can also develop disclination lines. However, this is beyond the scope of the present review, and we refer to Refs.~\cite{Fumeron2023,
Carenza2019Review,MerminRevModPhys,Afghah2018,duclos2020,Zushi2022} for a more detailed discussion on this topic.

The structure of topological defects inherently depends on the $p-$atic
symmetry of the system. In fact, a disclination can be characterised by looking at the configuration of the order parameter around its core and computing the winding number, or topological charge, $s$. This is a measure of the strength of the topological defect and is defined as the number of times, the order parameter turns of an angle of $2\pi$ while moving along a close contour surrounding the defect. 
The minimum defect charge (in module) available to a liquid crystal with $p-$atic symmetry is $s= \pm 1/p$, while defects with any charge $s= \pm n/p$ (with $n$ any integer) multiple of the fundamental one are topologically allowed. Elementary topological defects for nematic and hexatic liquid crystals are shown in Fig.~\ref{fig:3}(a-c) and Fig.~\ref{fig:3}(d-e), respectively.
Interestingly, the energy of a defective configuration can be derived from the Landau-de Gennes free energy in Eq.~\eqref{eq:5} and reads
\begin{equation}
\mathcal{F}^{\rm def} = \pi L_{p} s^2 \ln \left( \dfrac{R}{r_{\rm c}} \right), \,
\label{eq:6}
\end{equation}
where $r_{\rm c} \sim \mathcal{O}(\sqrt{L_{p}/A_{\rm bulk}})$ is a short-scale ultra-violet cutoff of the order of the coherence length of the liquid crystal, and $R$ is a large-scale infra-red cutoff of the order of the distance of the defect from the boundary, or other defects. 

Topological defects in $2D$ can be removed through a process of annihilation between defects of opposite charge. Indeed, one may consider a region populated by two oppositely charged defects, such as the ones shown in Fig.~\ref{fig:3}(c) and, by taking a contour which surrounds both defect cores, it is immediately found that the topological charge enclosed in the contour is null, thus signaling that such configuration can be continuously deformed into a perfectly aligned state. Generally, there is no rule which prevents either the formation or the elimination of disclinations, as long as the total topological charge of the system is conserved. This implies that the process of annihilation/nucleation of topological defects is a pair process. Analogously, a single topological defect of charge $s$ may also split into two topological defects with charges $s_{1}$ and $s_{2}$ provided that $s = s_{1} + s_{2}$. For instance a $+1$ vortex in a nematic liquid crystal may split into two two $+1/2$ defects, thus allowing for a reduction of the total energy
of the system. Indeed, the quadratic dependence on the defect charge in Eq.~\eqref{eq:6} plays an important role on the physical stability of defect configurations since a high-charged disclination pays a greater energy cost than a collection of defects which have the same total topological charge, but individually lower charge. However, this energetically favoured decaying process is topologically allowed only for liquid crystals with nematic symmetry. Conversely, polar LC can only develop defects with only integer charge, whereas semi-integer configurations are forbidden because of the lack of inversion symmetry.
Analogously, a $+1/2$ defect in a hexatic system is energetically favoured to split into three $+1/6$ defects, a process that is nevertheless forbidden for nematic defects as $+1/2$ is the minimum charge available to liquid crystals with nematic symmetry. 

In Section~\ref{sec:IV} we will further develop our analysis of topological defects in active systems with a more specific focus on their dynamics and relevance in biological and bio-inspired materials.

\subsection{Hydrodynamics of Active Fluids}
\label{sec:III-B}
Many biological and bio-inspired materials evolve in fluidic environments and exhibit themselves fluid-like characteristics, so that concepts such as flow, viscosity and stress, which are typical of the language of classical fluid dynamics, can be naturally extended to treat active systems. Analogously to passive liquid crystals, where the nematogens can be advected by external flows, active units can flow as a fluid, with the position of the self-propelling particles randomly distributed as the molecules in a fluid and their orientation locally ordered. 
The idea beyond the hydrodynamics description of active matter is to introduce a limited number of mesoscopic coarse-grained field able to retain the relevant information necessary to describe the state of the system, while tracing out the large amount of information regarding the individual dynamics of the active constituents. This continuous approach proved to be a valuable tool to describe many active systems, from bacterial colonies to epithelial tissues and cytoskeletal suspensions \cite{marchetti2013}.

In the following we will present the hydrodynamic equations for active matter in a $2D$ environment, focussing on the celebrated case of active nematics. 
The dynamical fields of the model are the density $\rho(\bm{r},t)$, the fluid velocity $\bm{v}(\bm{r},t)$ and the nematic tensor $\bm{Q}(\bm{r},t)$, and 
the dynamical evolution is ruled by the following set of partial differential equations
\begin{align}
\partial_t \rho + \nabla \cdot \left( \rho \bm{v} \right) = 0 \;, \label{eq:7} \\
\partial_t \bm{Q} + \bm{v} \cdot \bm{Q} - \left(\bm{Q} \cdot \bm{\omega} - \bm{\omega} \cdot \bm{Q} \right) = \lambda S \bm{u} + \gamma^{-1} \bm{H }  \;, \label{eq:8}  \\
\rho \left( \partial_t \bm{v} + \bm{v} \cdot \nabla \bm{v} \right) = -\nabla P + \eta \nabla^{2} \bm{v} + \nabla \cdot \bm{\sigma}^{\rm (e)} + \nabla \cdot \bm{\sigma}^{\rm (a)}  \;. \label{eq:9} 
\end{align}
Here, Eq.~\eqref{eq:7} is the continuity equation that enforces mass conservation. Since biological flows are usually weak, in many situations density fluctuations can be factually neglected. In this case, the continuity equation reduces to the solenoidal condition for the velocity field
\begin{equation}
\nabla \cdot \bm{v} = 0,
\label{eq:10}
\end{equation}
implying fluid incompressibility. 
The evolution of the $Q$-tensor is ruled by the Beris-Edwards equation in Eq.~\eqref{eq:8}, where the term on the left-hand side is the co-rotational derivatives and defines how the elongated constituents of the (active) liquid crystals are advected by the flow, being $\bm{\omega} = [\nabla \bm{v} - (\nabla \bm{v})^{T}]/2$ the vorticity tensor. The first term on the right-hand side is the flow-alignment contribution, with $\lambda$ the flow alignment parameter, $S$ the amplitude of the nematic order, as defined in the discussion following Eq.~\eqref{eq:2}, and $\bm{u}= [\nabla \bm{v} + (\nabla \bm{v})^{T}]/2$ the strain-rate tensor.
Finally, $\bm{H} = - \delta \mathcal{F}^{\rm LdG}/ \delta \bm{Q}$ is the molecular field which drives the relaxation of the liquid crystal towards its ground state defined by the Landau-de Gennes free energy. This was already introduced in Eq.~\eqref{eq:5} for the case of a generic $p-$atic liquid crystal for the complex order parameter $\Gamma_{p}$, and can be written in terms of the $Q-$tensor as follows $ \mathcal{F}^{\rm{LdG}} = \int {\rm{d}} {\bm{r}} \, [A_{\rm{bulk}}(-{\rm{Tr}}(\bm{Q}^2)/2 + {\rm{Tr}}(\bm{Q}^4)/4) + L (\nabla \bm{Q})^2/2]$. 
Finally, $\gamma$ is the rotational viscosity that determines the relevance of hydrodynamic advection with respect to elastic relaxation.

The dynamical evolution of the flow field is ruled by the Navier-Stokes equation in Eq.~\eqref{eq:9}. Here, $P$ is the hydrodynamic pressure and $\eta$ is the shear viscosity, while the stress tensor has been divided into two contributions.
The first is the elastic stress $\bm{\sigma}^{\rm (e)} = - L \nabla \bm{Q} : \nabla \bm{Q} - \lambda \bm{H} + \bm{Q}\cdot\bm{H} - \bm{H} \cdot \bm{Q}$ that can be derived from equilibrium considerations, while
the second term $\bm{\sigma}^{\rm (a)}$ is the active stress, a phenomenological contribution able to minimally capture the force exerted by the active nematic particles on the surrounding environment. For active nematics, this can be written as
\begin{equation}
\bm{\sigma}^{\rm (a)} = \alpha \bm{Q} \;,
\label{eq:11}
\end{equation}
where $\alpha$, the so-called activity, is a phenomenological parameter proportional to the intensity of the force exerted by the active constituents on the surrounding fluid.
The sign of the activity $\alpha$ is used to distinguish between puller or \emph{contractile} particles when $\alpha > 0$, and pusher or \emph{extensile} particles if $\alpha < 0$~\cite{hatwalne}.

Analogous hydrodynamics models for active liquid crystals of generic $p-$atic symmetry have also been developed, but their explicit treatment goes beyond the scope of this review.
We suggest the interested reader to refer to References~\cite{Negro2018,Giordano_2021,Carenza2019IntjC,Negro2019AIP} for polar liquid crystals, to References~\cite{doostmohammadi2018} for nematic symmetry, and to References~\cite{Krommydas2023,eLife86400} for hexatic symmetry. Moreover, a presentation of the hydrodynamics equations for liquid crystal with generic $p-$atic symmetry was recently developed in Ref.~\cite{Giomi2022patic,Krommydas2023}. 

Finally, we mention that models to confine (active) liquid crystals in arbitrary geometries, like a sphere or a spherical shell, have recently attracted much attention \cite{aronson2012,tjhung,tjhung2015,ruske2021,negro2025}. This is usually achieved by introducing a scalar field $\phi(\bm{r},t)$, that represents the concentration of active constituents and, for a spherical nematic droplet, through the following free-energy
\begin{equation}
\mathcal{F}^{\rm{drop}} = \int {\rm{d}} {\bm{r}} \, \left[ \dfrac{a}{4\phi_{0}^4} \phi^{2} (\phi - \phi_{0})^2 + \frac{k}{2}(\nabla\phi)^2+A_{\rm{bulk}} \left( - \dfrac{\phi - \frac{\phi_0}{2}}{\frac{\phi_0}{2}} \dfrac{{\rm{Tr}}(\bm{Q}^2)}{2} + \dfrac{{\rm{Tr}}(\bm{Q}^4)}{4} \right) + \dfrac{L}{2} (\nabla \bm{Q})^2 \right].
\label{eq:12}
\end{equation}
The first term is a Landau-Ginzburg potential for the concentration field, which has two minima in $\phi=0$ (the passive phase) and in $\phi=\phi_0$ (the phase rich of active fluid), while the second one is the energetic cost associated with fluid interfaces. 
Also, $a$ and $k$ are two positive constants controlling surface tension and interfacial width of the droplet. 
Confinement of the liquid crystal in the active phase is achieved by letting coefficient multiplying the quadratic term in the bulk liquid crystal energy depend on the concentration field. Therefore, this is negative in the region where $\phi > \phi_0/2$, and positive otherwise. 
More complex geometries (like spherical and toroidal shells) can be stabilized through similar approaches (see, for example, Ref.~\cite{Yeomans2019Hydra,Carenza2022prl,Negro2023softmatter} for more details). Furthermore, we suggest the interested reader to refer to Refs.~\cite{succi2018,denniston2001,sulaiman,negro2019_intc,Negro2024PRE} for numerical strategies to solve numerically the hydrodynamics of complex fluids and for their extension to active systems ~\cite{Carenza2019Review,Shaebani2019,tiribocchi2025}. 

Finally, we mention that, alongside theories including orientational order, over the last years there has been a growing interest towards scalar theories of orientational-free active fluids,~\cite{stenhammar2013,Tiribocchi2015,Negro_2019EPL,Wittowski2014,tjhung2018,cates2025,burekovic2026} which have been demonstrated to capture MIPS not depending on alignment interactions.

\section{Topological defects and Active Systems}
\label{sec:IV}
Topological defects play a crucial role in active biological systems, 
since they often serve as centers of rearrangement and morphogenesis, and even drive the dynamical properties of tissues, influencing the overall behavior and functionality of these systems.
In this Section we will explore further the dynamics of active topological defects, starting by 
a discussion about
how defects can determine the fate of living organisms, and then we will 
show how the active liquid crystal theory can be exploited to predict the motion of isolated topological defects. Finally, we will focus on the complex phenomenon of active turbulence, where activity is so large to provide enough energy to drive the unbinding of topological defects and lead to the onset of chaotic dynamics.  

\subsection{The biological role of topological defects}
\label{sec:IV-A}

\paragraph*{Defects in bacterial biofilm.} Bacteria can develop nematic phases while forming a biofilm. This feature is particularly notable in the case of biofilms of \emph{Bacillus subtilis}, which exhibit chaining behaviors (Fig.~\ref{fig:1}a-e) under specific environmental conditions~\cite{Yaman2019}. It was reported that activity-mediated mechanical instabilities, driven by localized buckling and edge instabilities, are essential in creating topological defects within the biofilm. These defects, observed as $\pm 1/2$ nematic disclinations, lead to morphological features, such as circular structures and localized folds, at lengthscales which directly depend on the bacterial interaction with other cells in the biofilm and with the outer environment, as for instance friction and adhesion forces between the bacterial chain and the surface.
In more detail, localised buckling of nematic biofilm fibers leads to the formation of $+1/2$ defects and in turn to an equal number of negative $-1/2$ disclinations to guarantee conservation of the total topological charge. Even more interestingly, the mechanics of these defects influence the biofilm's stress distribution, potentially impacting cellular physiology and behaviour. 

\paragraph*{Defects in epithelial tissues.} Another remarkable example of the biological relevance of topological defects is given by epithelial tissues. As we mentioned in Section~\ref{sec:III}, epithelial tissues feature the hallmarks of both nematic and hexatic liquid crystalline order and therefore exhibit topological defects of both $\pm 1/2$ and $\pm 1/6$ charge. Recent research suggests that both kinds of defects have their own functionality. 
In Ref.~\cite{Saw2017} Saw \emph{et al.} investigated the mechanisms underlying apoptotic cell extrusion in epithelial monolayers for an in vitro system of MDCK epithelial cells and observed that extrusion events were preceded by a coordinated flow of cells toward the extrusion site. These cell flows formed comet-like structures, identified as $+1/2$ topological defects, with the head of the comet pointing towards the cell destined for extrusion, finding a significant correlation between extrusion events and the positions of nematic disclinations. 

Furthermore, in Ref.~\cite{krommydas2023collective}, the relevance of hexatic topological defects was studied in a model epithelial tissue in the context of remodeling T1 events -- the process exploited by cells to rearrange and flow within confluent epithelial layers, even in the absence of gaps or interstitial structures. Here, it was proposed that collective epithelial migration can be interpreted as the unbinding of hexatic defect quadrupoles into free $\pm 1/6$ disclinations, therefore suggesting that hexatic defect distribution may play a central role in determining cellular migration in tissues.

\paragraph*{Hydra morphogenesis.} An even more striking example of the relevance of topological defects in biology is represented by the morphogenesis of organisms like the \emph{Hydra}, a small, tubular organism that lives in freshwater environments. The \emph{Hydra} is well-known for its remarkable regenerative abilities and its unique body structure, which is essentially organised into two layers: an outer layer, the epidermis, and an inner layer, the endodermis, both of which are separated by an intermediate gel-like layer. The body is radially symmetric, with a central mouth surrounded by tentacles.
In Ref.~\cite{MaroudasSacks2021} it has been showed that the \emph{Hydra} might exploit nematic topological defects in the organisation of actin fibres in the ectoderm to initiate the development of its morphological features, such as head, foot, and tentacles.  Indeed, the defects patterns, observed during the early stages of regeneration, allow for the identification of the sites where both the head and foot will develop, long before these morphological features become visible. 
Analogously, tentacle formation follows from the nucleation of topological defects of opposite charge from an initially defect-free region. More specifically, tentacles corresponds to $+1$ defects, while two $-1/2$ disclinations are systematically located at its base. 
These striking correlations strongly suggest that the defect distribution on the \emph{Hydra}'s body could  play a significant role in the establishment of the body plan, and fostered interesting theoretical interpretations which will be reviewed in more detail in the following.

\subsection{Dynamics of Active Defects}
\label{sec:IV-B}
One of the most important achievements of the active liquid crystal theory presented in Section~\ref{sec:III-B} is providing a framework for describing the dynamics of active topological defects, which proved particularly effective in the context of cytoskeletal suspensions.
The aim of this Section is to review the didactic case of the motion of an isolated active disclination, starting from the set of active hydrodynamic equations (Eqs.~\eqref{eq:7}-\eqref{eq:9} for an active nematic) and following Giomi's approach in Ref.~\cite{giomidefectsdynamics}.

\begin{figure}[t!]
\centering
\includegraphics[width=0.6\columnwidth]{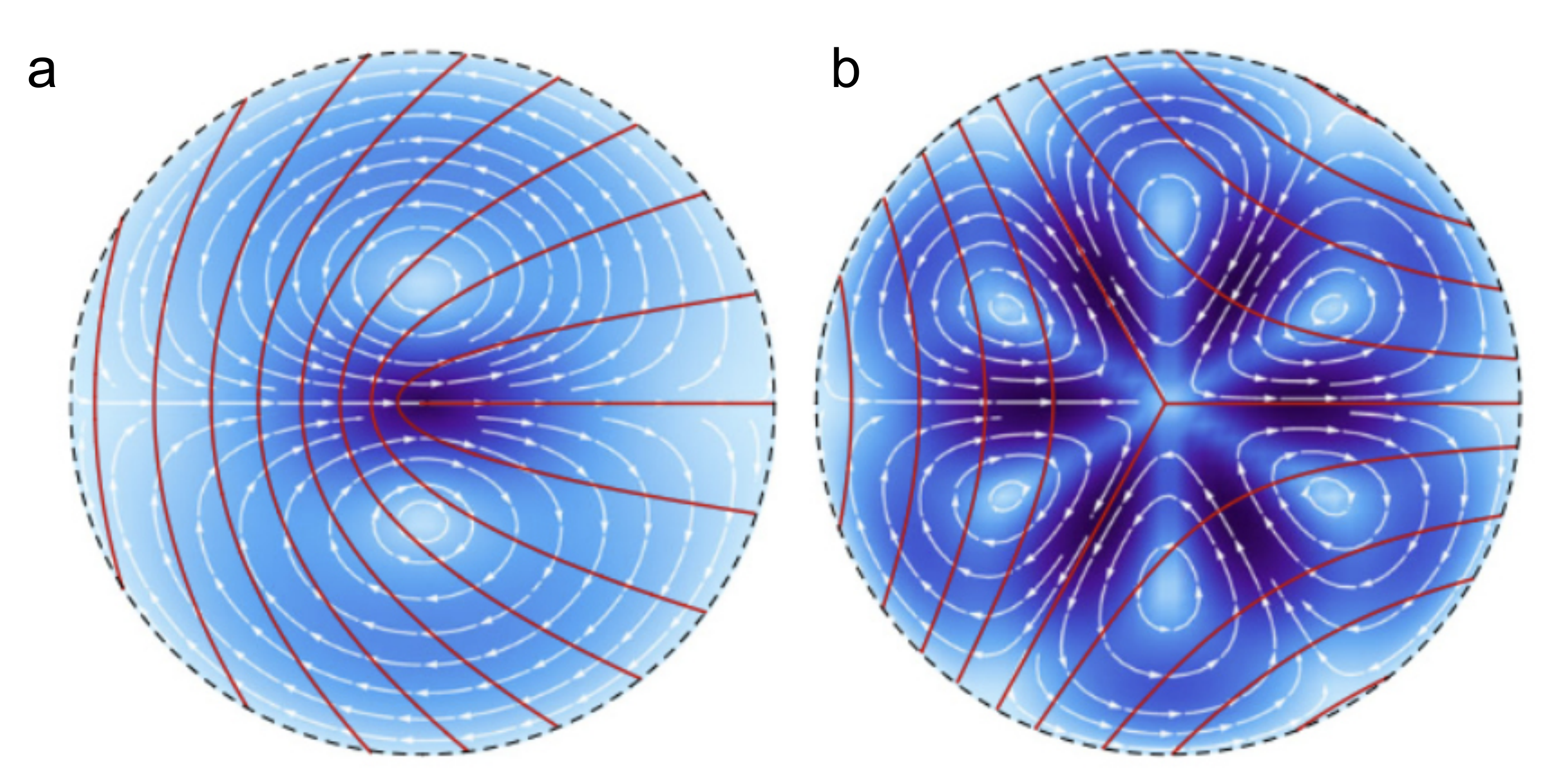}
 \caption{\textbf{Active flow of topological defects.} Active flow field (white streamlines)  of a $+1/2$ defect for contractile activity $\alpha>0$ (panel~a) and $-1/2$ for extensile activity $\alpha<0$ (panel~b). Red streamlines represent the orientation of the director field, and the background the local magnitude of the active flow, obtained by solving the Stokes equation, Eq.~\eqref{eq:13}, with no slip boundary conditions on a circle (dashed black line). Panels adapted from ~\cite{giomidefectsdynamics}.}
 \label{fig:flowdefects}
\end{figure}

In many biological realisations, the Reynolds number ${\rm Re} = \rho \bar{v} / \eta \ll 1$, where $\bar{v}$ represents the typical speed of active flows. This means that advective and inertial effects are negligible with respect to viscous dissipation. Under these conditions, the Navier-Stokes equation can be significantly simplified into the Stokes equation
\begin{equation}
\eta \nabla^{2} \bm{v} - \nabla P + \nabla \cdot \bm{\sigma}^{\rm (e)} + \nabla \cdot \bm{\sigma}^{\rm (a)} = 0 \;,
\label{eq:13}
\end{equation}
with $\bm{v}$ the solenoidal velocity field ($\nabla \cdot \bm{v} = 0$).  Considering isolated disclinations of charge $\pm 1/2$ located at the centre of a circular domain of radius $R$,  by virtue of linearity the flow solution of the Stokes equation can be expressed as
\begin{equation}
\bm{v} = \bm{v}^{0} + \bm{v}^{a} + \bm{v}^e \;,
\label{eq:14}
\end{equation}
where $\bm{v}^{0}$ is the solution of the homogeneous Stokes equation, and $\bm{v}^{a}$ and $\bm{v}^{e}$ are the flow field sourced by the active and elastic stress contributions, respectively.
Making use of the Oseen solution of the $2-$dimensional Stokes equation, the explicit expression of the (active) flow can be computed as
\begin{equation}
\bm{v}(\bm{r}) = \int {\rm{d}} {\bm{r}'} \; \bm{G}(\bm{r}-\bm{r}') \cdot \bm{f}^{(\rm a)} \;,
\label{eq:15}
\end{equation}
where $\bm{G}(\bm{r})$ is the $2-$dimensional Oseen tensor~\cite{Krommydas2023} and $\bm{\sigma}^{(\rm a)}$ is the force density  $\bm{f}^{(\rm a)} = \nabla \cdot \bm{\sigma}^{(\rm a)}$ given by the divergence of the active stress tensor in Eq.~\eqref{eq:11}, whose form can be approximated using the equilibrium configuration of the $Q$ tensor $\bm{Q} \sim (\bm{n}\bm{n}-\bm{I}/2)$, with $\bm{n}= \cos(s\phi) \hat{\bm{x}} + \sin(s\phi) \hat{\bm{y}}$ the director field in proximity of a topological defect of charge $s$.
Figure~\ref{fig:flowdefects} show the graphical solution of the active flow field for a positive (panel~a) and negative (panel~b) nematic disclination~\footnote{The analytical solution of the active flow generated by a positive/negative semi-integer active nematic defect is given by Eq.~(3.9)a,b in Ref.~\cite{giomidefectsdynamics}.}.
Importantly, while negative defects correspond to a stagnation flow field with no mass transfer, positive defects are able to generate a directed flow, which leads to the self-propulsion of the active defect.
Moreover, the extensile or contractile nature of the active fluids plays a critical role in determining the direction of propulsion of a $+1/2$ defect. Indeed, contractile ($\alpha>0$) defects generate a net flow along the symmetry axis of the disclination that is directed from the head towards the tail of the comet (Fig.~\ref{fig:flowdefects}a)  whereas, in extensile systems ($\alpha>0$), the flow is reversed and the direction of self-propulsion is directed towards the head (Fig.~\ref{fig:flowdefects}b). 

The self-propulsion feature of nematic active defects drastically affects the steady-state properties of active liquid crystals and the mutual interaction between defects. Indeed, while in equilibrium systems oppositely charged defects annihilate under the effect of the elastic attraction, in active systems annihilation can either be fastened or prevented under the fueling action of activity, in accordance to the relative orientation of defects of opposite charge.
Moreover, under high levels of activity, defects pairs can also be produced starting from a defect-free region, and unbind. This phenomenon, observed in various active systems from microtubule and actomyosin suspensions to bacterial swarming, results in the proliferation of topological defects, which continuously nucleate and annihilate, leading to a condition of dynamical equilibrium. This fascinating chaotic regime is known in the literature with the name of \emph{active turbulence} and it will be the focus of the following Section.

\subsection{Active Turbulence}
\label{sec:IV-C}

\begin{figure}[t!]
\centering
\includegraphics[width=0.95\columnwidth]{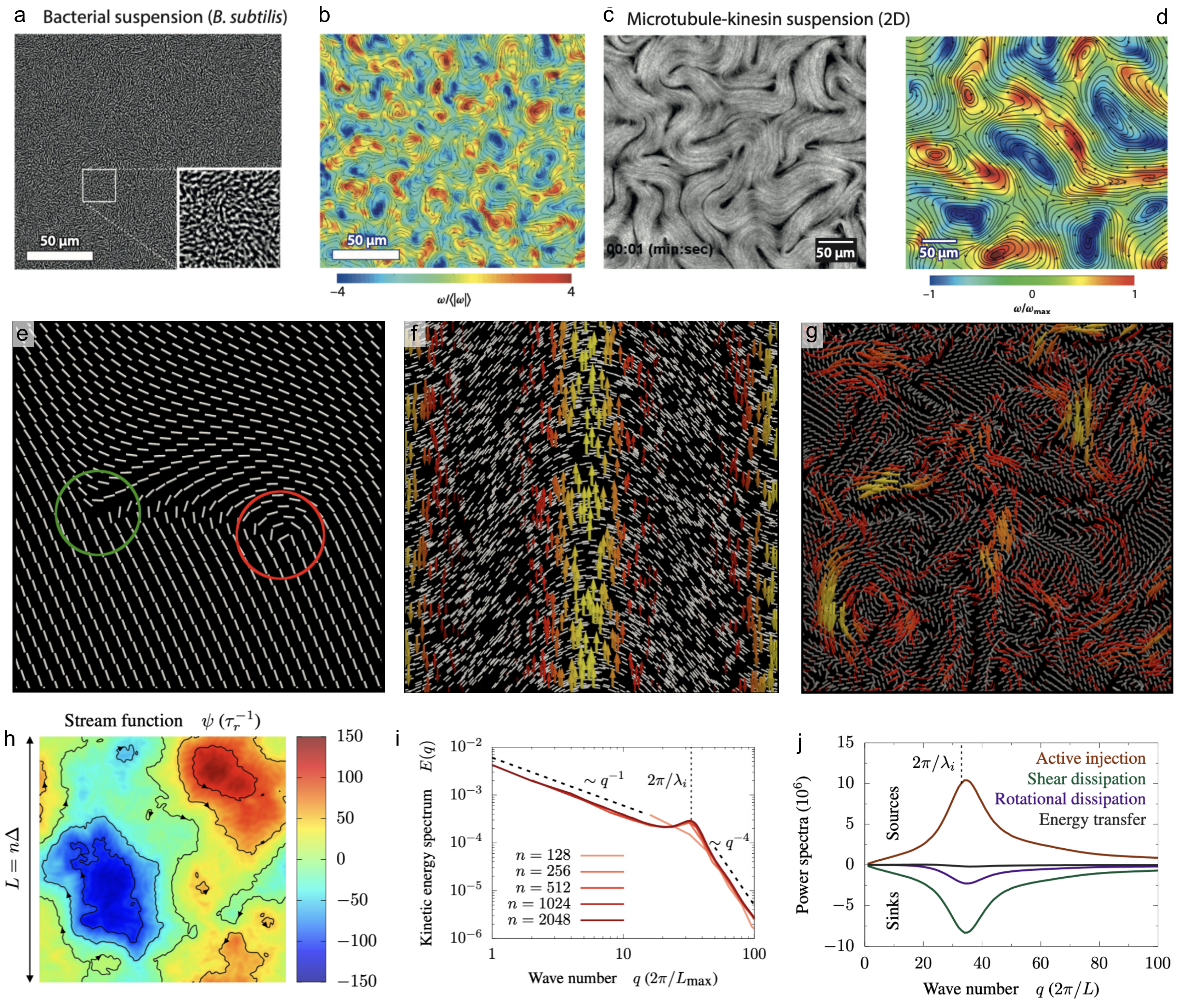}
 \caption{\textbf{Active Turbulence} (a) Experimental image of a B. subtilis suspension. The corresponding reconstructed vorticity field, with streamlines, is shown in (b). Adapted from Ref. \cite{Wensink2012}. (c-d) Experimental image and vorticity field of a microtubule-kinesin suspension. Adapted from Ref. \cite{Guillamat2017}. (e-g) Numerical simulations of an active nematics at varying the intensity of the activity. The picture is a summary of the three dynamical regimes occurring in active gels as reported in~\cite{giomidefectsdynamics,Thampi_2014,Giomi2011PRL,Urzay_Doostmohammadi_Yeomans_2017}. Panel (e) shows the configuration of the nematic field (white rods) during the relaxation dynamics in proximity of two oppositely charged semi-integer defects in the quiescent regime. Increasing activity, the system shows the emergence of a spontaneous flow regime (f) where the nematic field undergoes an activity-induced banding instability that breaks the rotational symmetry, producing bands of laminar flow (colored arrows), stronger where elastic deformations are more pronounced. At even higher activity (g), a chaotic regime is encountered. This state is characterized by the formation of walls, narrow regions of strong banding deformations, which eventually give rise to the enucleation of topological defects. Panels (e)-(f)-(g) adapted from Ref. \cite{carenza2020_bif} .
 (h) Typical pattern of the stream function field.  The small background ripples correspond to coherent vortices of typical size given by the scale of maximal energy injection. However, large-scale correlated flow patches also form owing to
the non-local character of viscous flow. Lines with arrows indicate a few streamlines that highlight these large-scale circulations. (i) Spectra of kinetic energy exhibit a peak at the maximal injection scale. At smaller and, importantly, at larger scales, the spectra feature distinctive universal scaling regimes. 
(j) Spectra of the contributions to the energy balance in simulations of a minimal model of wet active nematics. There is no energy transfer across scales; the energy injected at each scale is dissipated at that same scale by the combined effects of shear and rotational viscosity. Panels h,i,j adapted from Ref. \cite{Alert2020}.
 }
 \label{fig:4}
\end{figure}
Active systems often exhibit a chaotic dynamics. This behaviour, first observed in bacterial systems \cite{Dombrowski2004}, is characterized by swirling patterns in the organization of the active units, reminiscent of those observed in a classical isotropic turbulent fluid flowing at high Reynolds number. Due to such an analogy, the chaotic state was initially called \emph{bacterial turbulence} and then turned into \emph{active turbulence}~\cite{Alert2022}, to characterise analogous chaotic behaviours observed in a plethora of other realisations, from cytoskeletal suspensions~\cite{Guillamat2017} to bacterial~\cite{Peng2021} and sperm swarming~\cite{Creppy2015}, cellular tissues~\cite{Lin2021} and colloids~\cite{Nishiguchi2015}. Panels (a-d) in Fig.~\ref{fig:4} shows examples of the typical  chaotic patterns  observed in these systems. 
Since the first observation, much effort has been spent in studying and characterising the chaotic behaviour of active systems, starting from the appropriateness of the term \emph{turbulence} commonly used to describe it, which, far from being semantic, concerns the nature of the mechanism driving the dynamics of the system. Indeed, turbulence in a classical isotropic fluid occurs at large Reynolds numbers as the effect of the energy cascade due to nonlinear advection from the scales at which it is injected to those where it is dissipated~\cite{boffetta2012}. Conversely, biological fluids mostly evolve at vanishing or very low Reynolds numbers, where nonlinear effects can be factually neglected. 

Much theoretical and computational research has focused on the case of active turbulence in nematics~\cite{doostmohammadi2018,Alert2022,Head2024}. In this framework, the transition to the \emph{turbulent} regime is driven by the interplay between active injection and the elastic relaxation arising from liquid crystal alignment. In more detail, three regimes can be typically observed in active nematics. At low activity, a \emph{quiescent} regime is found where the liquid crystal actually behaves as a passive system  (Fig.~\ref{fig:4}e). At intermediate active forcing, the liquid crystal becomes unstable and develops banded features sustained by activity-induced \emph{spontaneous flow}, counteracted by elastic effects (Fig.~\ref{fig:4}f). Finally, as activity is further increased and active flow grows, liquid crystal deformations become progressively sharper, eventually favoring the nucleation of topological defect pairs of charge $\pm 1/2$ (Fig.~\ref{fig:4}g). The self-propulsion properties of positive disclinations~\cite{giomidefectsdynamics} further enhance the production of new defect pairs, disrupting the residual order and restoring isotropic symmetry at a statistical level. This process of continuous defect nucleation and annihilation lead to a non-equilibrium steady state resulting in what is called fully developed active turbulence \cite{bonn2026}.

A mean-field theory able to describe the statistical features of the turbulent active flows was first proposed by Giomi~\cite{Giomi2015},  focusing on the statistical properties of the vortical patterns (highlighted in Fig.~\ref{fig:4}h). Assuming an exponential distribution for the vortex area, it is showed that turbulent active flows could develop energy spectra, $E_q$, featuring power-low decay $E_q \sim q^{-4}$ at the small scales, and $E_q \sim q^{-1}$ at the large scales (see Fig.~\ref{fig:4}i). More recently~\cite{Alert2020}, it was also demonstrated that the large-scale scaling regime $q^{-1}$ is universal, independent of the material parameters of the active fluids (such as sign and magnitude of activity, viscosity, elasticity, and flow alignment) and of the presence of topological defects.

Remarkably, the existence of (universal) scale-invariant features of the energy spectrum is once more reminiscent of hydrodynamic turbulence. In this case, the energy spectrum follows a power law Kolmogorov decay $E_q \sim q^{-5/3}$, inherently related to the transfer of energy across scales, due to advective effects.
While it is clear that advection cannot be the driving mechanism in active nematics (and other biological and bio-inspired realizations), other non-linear interactions at the level of the active Navier-Stokes equation could still be, in principle, responsible of an energy cascade, hence of the observed scale-invariant universal features of the active energy spectrum $E_q$. To this aim, a balance equation for $E_q$ can be derived starting from the Navier-Stokes equaction~\cite{Alert2020,Carenza2020,carenza2020_bif} 
\begin{equation}
\partial_t E_q + \mathcal{T}_q = \mathcal{S}^{\rm{(v)}}_q + \mathcal{S}^{\rm{(e)}}_q + \mathcal{S}^{\rm{(a)}}_q,
\label{eq:16}
\end{equation}
where $E_q=\langle  | \bm{v}_{\bm{q}}|^2 \rangle/2$ is the energy spectrum,  with $\bm{v}_{\bm{q}}$ the Fourier representation of the flow field and $\langle \cdot \rangle$ denoting the sum on shells of equal momentum ($|\bm{q}|=q$), and
$\mathcal{T}_{q}=  \langle \bm{v}_{\bm{q}}^* \cdot \bm{J}_{\bm{q}} \rangle$ is the rate at which energy is transferred by non-linear advective interactions, with $\bm{J}_{\bm{q}}$ the Fourier representation of the hydrodynamic flux $-\nabla p + \bm{v}\cdot \nabla \bm{v}$.
The terms on the right-hand side of Eq.~(\ref{eq:16}) represent the rates at which viscosity removes energy from the system, $\mathcal{S}^{\rm{(v)}}_q \propto - \eta q^2 E_q$, and the rate at which elastic $\mathcal{S}^{\rm{(e)}}_q$ and active $\mathcal{S}^{\rm{(a)}}_q$ contributions absorb, inject and eventually transfer energy among length-scales. 
Numerical studies demonstrated that at steady-state ($\partial_t E_q=0$), hydrodynamic non-linearities are, at all practical effects, negligible at any scale ($\mathcal{T}_q \approx 0$), as expected in a system evolving at low Reynolds numbers. 
Furthermore, while in Refs.~\cite{carenza2020_bif,Alert2020} elastic energy transfer was reported, this would hardly be responsible for an energy cascade,  being the dynamics mainly dominated by the scale-by-scale balancing between active injection and viscous dissipation ($\mathcal{S}^{\rm{(a)}}_q \approx - \mathcal{S}^{\rm{(v)}}_q$ at any wavenumber $q$).

In conclusion, active turbulence is an inherent feature of many active systems and directly follows from the complex interplay between the activity-mediated energy injection and the dissipative mechanisms. While a broad understanding of this phenomenon has been achieved in the past two decades, further questions regarding the relevance of friction, chirality, and geometry still demand further investigation. In particular, a challenging objective of both theoretical and experimental investigation is to develop a strategy to tame active turbulence, a goal which would allow for the macroscopic control of active systems. In the next Section we will review how geometrical confinement, topological constraints and external fields can provide an effective approach to regulate the transition to the turbulent regime. 

\section{Geometrical confinement,  topological constraints and external fields}
\label{sec:V}

\begin{figure}[t!]
\centering
\includegraphics[width=0.95\columnwidth]{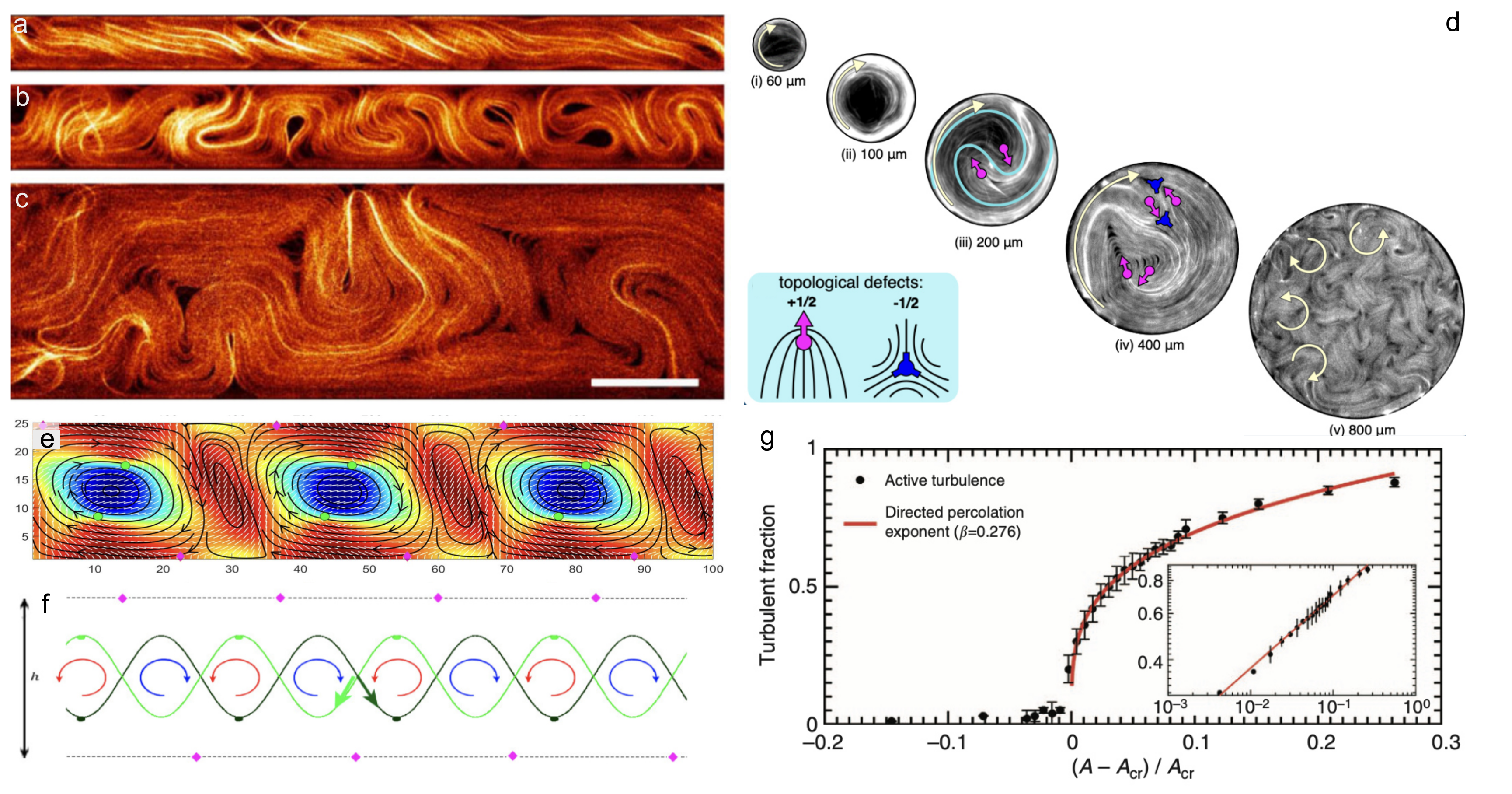}
 \caption{\textbf{Geometrical confinement} (a–c) Confocal fluorescence micrographs of an active nematic interface confined in channels of different widths. Scale bar:  $ 100 \mu {\rm m}$. The colour map corresponds to the computed nematic order parameter. Adapted from Ref.~\cite{Confnematics1}. (d) Fluorescence images of  confined active nematics in circular cells with diameter ranging from $60$ to $800 \mu m$- Adapted from Ref.~\cite{Achini2019}. Yellow arrows indicate direction of circulation. Blue line overlaying $200\mu m $  highlights the double spiral configuration of the nematic director observed for intermediate confinements. (e) Simulation snapshot of the instantaneous vorticity field, streamlines and director field. The vorticity field (colored background) forms an ordered vortex lattice and a dynamically ordered state of disclinations within the channel. Positive ($+1/2$) and negative ($-1/2$) disclinations are denoted as green dots and magenta diamonds, respectively. (f) Schematic of ideal Ceilidh dynamics. Negative $-1/2$ disclinations (magenta diamonds) are located in proximity of the walls and oscillate about their average position. In the center of the channel a vortex lattice of counter-rotating neighbouring vortices is found (red and blue curved arrows) with the $+1/2$ disclinations (green lines) moving along the boundary of the vortex lattice. Panels (e) and (f) adapted from Ref.~\cite{Shendruck2017}. 
 (g) Turbulence fraction as a function of the activity number. The red line is a power law with exponent $\beta=0.276$ for $1+1$ directed percolation with spontaneous activation. Adapted from Ref.~\cite{Doostmohammadi2017}.}
 \label{fig:5}
\end{figure}

In the previous Section we discussed how active systems exhibit a natural tendency to develop chaotic dynamics, making the objective of controlling and directing active flows an arduous task. Here we will discuss three possible strategies that can be implemented to control active systems, through geometry, topology and external fields.

\subsection{Active nematics in a channel} 
As a paradigmatic example, we consider the confinement of active nematics in prescribed geometries, which allows for a direct control over the spontaneous flow instability~\cite{Hardouin2022,Confnematics1,Doostmohammadi2019}. 
Experiments on active nematics in a channel~\cite{Confnematics1,Doostmohammadi2019} have shown that different  defect landscape and flow patterns can be obtained by changing the confinement strength, namely the height of the channel (Fig.~\ref{fig:5}a-c).  In general, two dynamical regimes can be identified: a shear flow regime, observed for low confinement (see Fig.~\ref{fig:5}a), which is transiently defect-free, and a dancing defects regime for higher channel height (see Fig.~\ref{fig:5}b). Increasing the channel height further affects the direction of the shear, which is 
composed of patchy domains where shear flow spontaneously arises with a random direction and is then quickly disrupted by instabilities, akin to an active turbulence regime (see Fig.~\ref{fig:5}c).  Confinements in disks also shows analogous dynamical regimes at varying the confinement strength (Fig.~\ref{fig:5}d)~\cite{Achini2019}. In small disks with radii comparable to the length-scale of active instabilities, topological defects, continuously nucleating and annihilating, self-organize into persistent circular flows of either handedness whereas, in larger disks, the dynamics becomes progressively turbulent.

It is instructive to study explicitly the onset of the instability  in a channel with a linear stability calculation. To this aim, we consider a quasi-$1D$ geometry, with the active nematics confined between two plates, perpendicular to the $x$ axis, and located at positions $x=0$, and $x=\mathcal{L}_x$. The boundary conditions are
\begin{align}
\bm{v}(x=0,\mathcal{L}_x)=0 \;, \label{eq:17} \\
\bm{Q}(x=0,\mathcal{L}_x) = S_0 \left( \bm{n}_0\bm{n}_0 - \bm{I}/2 \right)  \;, \label{eq:18}
\end{align}
with $\bm{n}_0=(1,0)$ the unit vector perpendicular to the walls and $S_0$ the equilibrium amplitude of nematic order. Uniformity in the channel direction $y$ implies that all quantities only depend on the $x$ coordinate. Furthermore, from the continuity equation in the incompressible limit 
one immediately obtains that no flow can occur transversely to the channel and the only non-null component of the velocity field is $v_y$. 
We now consider the stability of the trivial solution $\bm{v}_0=0$ and $\bm{Q}_0=S_0 \left( \bm{n}_0\bm{n}_0 - \bm{I}/2 \right)$ to infinitesimal perturbations of this reference state.
In particular, we choose $\bm{n}=\bm{n}_0 + \delta n_y(x) \hat{\bm{y}}$ and the velocity field $\bm{v}= \delta v_y(x) \hat{\bm{y}}$.
By substituting these in Eq.~\eqref{eq:8} and~\eqref{eq:9}, the linearised problem in terms of the perturbed fields reads as follows
\begin{align}
\partial_t \delta v_y = -\partial_y P + \eta \partial^2_x \delta v_y + \partial_x(\sigma^{\rm{(e)}}_{yx}+\sigma^{\rm{(a)}}_{yx}) \;, \label{eq:19}\\
\partial_t  \delta n_y = \dfrac{\partial_x \delta v_y}{2} + \dfrac{L}{\gamma} \partial_x^2 \delta n_y + \mathcal{O}(\delta n_y^2)  \;, \label{eqn:20}
\end{align}
where we set the flow-alignment parameter to $0$ for the sake of simplicity and the linearised elastic and active stress tensors are $\sigma^{\rm{(e)}}_{yx}= -4L \partial^2_x \delta n_y$, and $\sigma^{\rm{(a)}}_{yx}= 2 \alpha  \partial_x \delta n_y$, respectively. 
Taking twice the curl of Eq.~\eqref{eq:19} to eliminate the pressure, that equation becomes
\begin{equation}
\partial_t \partial_x^2 \delta v_y = \eta \partial_x^4 \delta v_y - 4L \partial_x^5 \delta n_y + 2\alpha  \partial_x^3 \delta n_y.
\label{eq:21}
\end{equation}
Then, by introducing the Fourier representation $\tilde{f}(\bm{q},\omega)$ of a generic field $f(\mathbf{r},t) = \sum_q \int  {\rm d} \omega \, \tilde{f} e^{ikx + \omega t}$, with $q=2 \pi n /\mathcal{L}_x$ the wave number of the $n$-th Fourier mode, and substituting in Eq.~\eqref{eq:19} and~\eqref{eq:21}, one obtains the evolution equation for each mode in Fourier space
\begin{align}
\omega \delta \tilde{v}_y = -\eta q^2 \tilde{v}_y + i (4L q^3 + 2 \alpha q) \delta \tilde{n}_y \; , \label{eq:22}\\
\omega \delta \tilde{n}_y =  \dfrac{i}{2} q \delta \tilde{v}_y - \dfrac{L}{\gamma}  q^2 \delta \tilde{n}_y \; , \label{eq:23}
\end{align}
which is an algebraic system of two equations for the two unknowns. Solving for $\omega$, yields the following dispersion relation
\begin{equation}
\omega_{\pm}(q) = -\dfrac{\eta+ \gamma^{-1}L}{2}q^2 \pm \dfrac{1}{2}
\sqrt{ [(\eta - \gamma^{-1} L)^2 - 8 L ^2] q^4  - 4 q^2 \alpha} \; .
\label{eq:24}
\end{equation}
In the passive limit $\alpha=0$, both branches gives rise to $Re[\omega_\pm]<0$, corresponding to diffusive modes ($\sim q^2$) so that perturbations spontaneously extinguish. For contractile activity ($\alpha > 0$), the solution may eventually develop a non-null imaginary part $Im[\omega_\pm]$ corresponding to acoustic modes (i.e. propagating waves) for large enough values of $\alpha$. However, these would eventually be damped anyway by the elastic and viscous diffusive modes. The most interesting case is given by the extensile case, where $\alpha < 0$ and can therefore make the branch $\omega_+$ real and positive.  This first occurs when the activity $\alpha$ exceeds the critical threshold
\begin{equation}
|\alpha_{cr}| = - \dfrac{ 4 \pi^2}{\mathcal{L}_x^2} \left( \dfrac{\eta L}{\gamma} + 2 L \right).
\label{eq:25}
\end{equation}
The resulting state is characterized by a sinusoidal pattern of the active liquid crystal, which gives rise to a self-sustained transverse  flow, with two lanes of fluid moving in opposite directions following the concavity of the liquid crystal, compatibly with the pattern shown in Fig.~\ref{fig:4}f. As activity increases, the modes with higher momentum are excited, with the consequent development of more lanes. Interestingly, from Eq.~\eqref{eq:25} one can also observe that, for a fixed value of the activity parameter, there is a critical channel width at which the active liquid crystal undergoes the transition towards a spontaneous flowing regime. Instead, channels thinner than this critical width stabilise the aligned equilibrium pattern.

This didactic presentation follows from many theoretical and computational studies where the behaviour of active systems confined in channels has been thoroughly eviscerated and many applications have been proposed, including the possibility of designing appropriate logic port to perform computational operations \cite{kos2022,zhang2022}.
Furthermore, simulations of the active turbulence regime in a channel presents interesting features that are worth mentioning here. As activity is increased beyond the threshold of the spontaneous flow instability, cyclic periodic steady states can be reached. This is the case of the \emph{Ceilidh dance}~\cite{Shendruck2017} (see Fig.~\ref{fig:5}e-f), a dynamical regime that, metaphorically, refers to the collective regular motion of topological defects in conjunction with a definite vortex lattice. Just as participants in a Ceilidh dance move in coordinated patterns, pairs of disclinations move along the channel and continuously switch neighbours, producing a dynamic ordered state. This periodical dynamical state is a precursor of the transition of the fully developed active turbulence -- a process which, in a channel, belongs to the directed percolation universality class (Fig.~\ref{fig:5}g).

Finally, we also mention that confined active fluids feature interesting rheological properties, such as the occurrence of states with apparent negative viscosity found in confined bacterial suspension~\cite{Santillan2018,Liu2019,Lopez2015}, and rationalized in the context of the active liquid crystal theory~\cite{Negro2019rheo,hatwalne}, but also active shear thinning and thickening and multi-stable rheological dynamics~\cite{Carenza2020rheo,Favuzzi2021}.  The rheology of biological tissue also plays an important role in many processes, from organ formation to cancer invasion~\cite{Hopkins2022,Manning2021}. 

\subsection{Topology, morphogenesis and self-motility} 
\begin{figure}[t!]
\centering
\includegraphics[width=1.0\columnwidth]{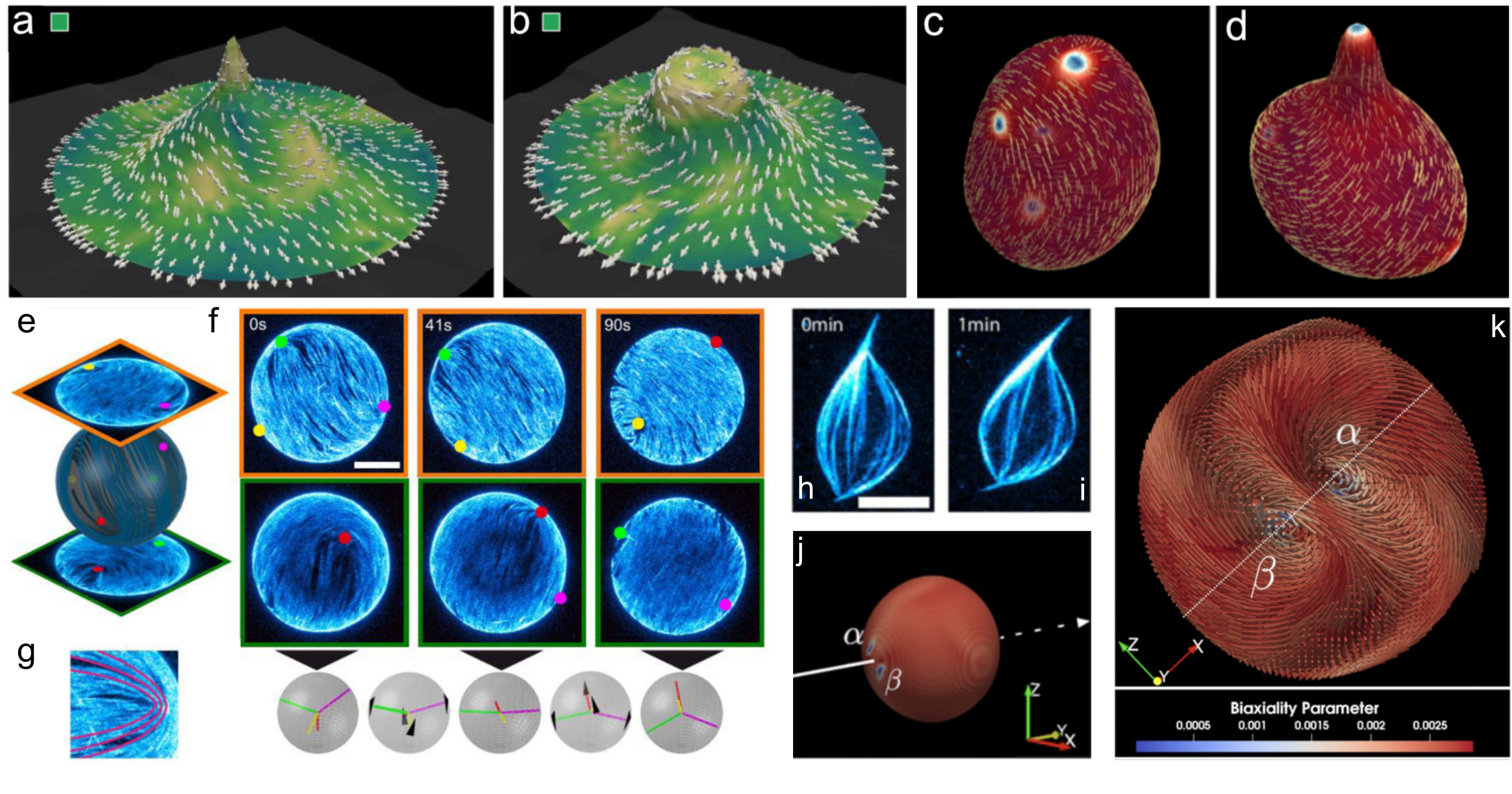}
 \caption{\textbf{Topological confinement}(a-d) Defect-mediated buckling and activity-driven shape deformations  obtained from numerical simulations of an active polar layer (a-b), and and active nematic shell (c-d).  The white vectors in (a-b) mark the polarization field, while the  yellow rods in (c-d) represent the director field. 
 In the case of an active polar layer, 
 beyond the threshold of the buckling instability, the layer attains a funnel-like profile with the topological defect at the end of the hollow (a).
 At slightly higher activity,  the interface begins to oscillates between this funnel-like state and the configuration shown in (b). Panels (a) and (b) adapted from Ref.~\cite{Hoffmann2022}. In the case of an active nematic shell, for sufficiently  large extensile activity, defects (hilighted by blue regions) deform the sphere periodically (c) and, for sufficiently large activity, protrusions are formed  as shown in (d). Panels (c) and (d) adapted from Ref.~\cite{Hoffmann2023}.  (e) Hemisphere projection of a 3D confocal stack of a nematic vesicle. The positions of four $+1/2$ disclination defects are identified. (f) Time series of hemisphere projections over a single period of oscillation in which the four defects switch from tetrahedral ($t = 0 s$) through planar ($t = 41 s$) and back to tetrahedral ($t = 90 s$) configurations. (g) Comet-like $+1/2$ disclination defect with schematic of the orientation of the nematic director (red lines). 3D Defect configurations  are shown in the bottom row of panel (f). The black arrowheads indicate the direction of defect motion.  (h-i) Spindle-like vesicle. Each pole contains a +1 aster-like defect.
 Panels (e,f,g,h,i) adapted from Ref.~\cite{Keber2014}. (j-k)  Snapshots at different times of a chiral active droplet. The droplet performers an uni-directed motion (dashed arrow in panel (j)). The contour plot of the biaxiality parameter on the droplet surface serves to identify the position of the $2$  defects labeled with Greek letters $\alpha$ and $\beta$, the configuration of which can be appreciated by looking at panel (j). Panels (j) and (k) adapted from Ref. ~\cite{CarenzaPnas2019}.}
 \label{fig:6}
\end{figure}
A further route to control topological defects is through 
topological constraints that may guide their stabilization (an objective of particular relevance even in passive liquid crystals \cite{RojoGonzalez2024}). 
In this context, a classical result of algebraic topology which goes under the name of Gauss–Bonnet theorem provides a natural starting point.
For the $3D$ physical space,  the theorem states that: 
\begin{quote}
\emph{Given a vector field on a $2D$ (curved) surface embedded in $3D$, with $N$ topological defects of charge $s_i$, with $i=1,\dots,N$, then 
\begin{equation}
\sum_{i=1}^{N} s_i = 2 - 2g - b \equiv \chi \;, 
\label{eq:26}
\end{equation}
where $\sum_{i=1}^{N} s_i$ is the total topological charge of the vector field, $\chi=2 - 2g - b$ is the Euler characteristic of the surface, $g$ is the number of holes in the surface and $b$ the number of boundaries.}
\end{quote}
Irrespective of the $p-$atic degree of orientational order, and as long as the orientational order field is confined on the surface, this will have to adapt to the topology of the surface itself. An example is the case of a sphere (or any other surface homotopic to a sphere) that does not have either holes or boundaries, so that $\chi=2$. Any vector field confined on the surface of a spheroid will therefore develop topological defects for a total topological charge equal to $+2$. This could be as a single $+2$ defect, two $+1$ defects, four $+1/2$ defects for a field with nematic symmetry, twelve $+1/6$ defects for hexatics and so on. Moreover, production of further pairs of oppositely charged defects is also \emph{topologically} allowed, as long as the condition in Eq.~\eqref{eq:26} is respected. Other interesting cases are disk-like geometries with $\chi=1$ ($g=0$, $b=1$) and toroidal surfaces with $\chi=0$ ($g=1$, $b=0$).

These results have very relevant biological implications. For instance, in Section~\ref{sec:III-A} we have discussed how the \emph{Hydra} exploits the existence of topological defects to develop morphological features and that disclination patterns become visible much before of the development of the morphological features. Interestingly enough, the body of the \emph{Hydra} has indeed spherical topology, so that formation of defects of total $+2$ topological charge is indeed required. Defects appear in the early stage development as two $+1$ disclinations to comply to the Poincar\'e-Hopf theorem, and will become eventually the mouth and the foot of the animal.

Recent research has suggested that the stabilization of  defects through topological confinement could indeed operate in coordination with activity to produce complex geometrical shapes and even lead to topological transitions~\cite{Hoffmann2023,negro2025,Yeomans2023PRXlife,Yeomans2019Hydra,Ruscke2021}. 
A series of studies~\cite{Hoffmann2022,Vafa2022,Fridtjof2023,Hoffmann2023} has demonstrated, for example, that defective active layers with liquid crystal order are unstable to out-of-plane deformations whose instability primarily depends on the orientational stiffness, surface tension, and strength of internal active stresses. Fig.~\ref{fig:6}(a-b) shows the numerical simulation of a disk-like active membrane with polar order embedded in $3D$ where the  boundary conditions for the liquid crystal stabilize a $+1$ topological defect at the center, in agreement with the Gauss-Bonnet theory. Here, activity generates a vortical flow that drives the buckling instability of the membrane which attains a funnel-like shape with the topological defect located at the tip of the protrusion. 

Another interesting case of study regards active shells of microtubules assembled at the interface of water-in-oil droplets. In a study by Keber \emph{et al.}~\cite{Keber2014} it was observed that the four nematic $+1/2$ disclinations could develop a periodical motion, fueled by activity, from the equilibrium tetrahedral configuration to a planar one 
(Fig.~\ref{fig:6}e-i),
with the vesicle undergoing large deformations and developing long protrusions. Interestingly, here nematic defects have been found to merge in two pairs of $+1$ dislocations  located in correspondence of the vescicle protrusions. These experimental observations are well reproduced by the active liquid crystal theory. Indeed, confining an active nematic liquid crystal on a deformable spherical shell recovers the periodical motion of defects (Fig.\ref{fig:6}c) at intermediate values of activity, while at larger values nematic defects can merge giving rise to an integer defect which in turn drives pronounced protrusions (Fig.\ref{fig:6}d) analogue to those observed for a flat membrane.
The dynamics of the active defects has been also exploited in experiments where
point-like active disclinations synchronize with  deformations of passive disclination lines 
in active nematic shells embedded in a passive 
liquid crystalline medium
\cite{Guillamat2017}.

Finally, we note that topology can also play an important role in the dynamics of dense droplets both in two- and three-dimensions. 
We mention here, for example,  the results observed for an active \emph{chiral} droplet 
\cite{CarenzaPnas2019,CARENZA2020PhysicaA} where
the chirality represents the tendency of a liquid crystal to develop an helical structure.
This leads to 
a structural reorganization of the surface defects, such as in the Frank-Price structure where two $+1$ defects merge to form a single $+2$ defects~\footnote{The final structure is akin to the topological state predicted for a magnetic monopole with its attached Dirac string, joining the centre of the droplet with the
$+2$ defect pair on the surface.}. When fuelled by activity, the droplets starts rotating around the axis defined by the surface defect, and the internal chiral structure converts the rotational motion into a translational motion – thereby resulting in a screw-like dynamics.
The relevance of topology and geometry in the evolution of biological and bio-inspired materials has been only partially explored, yet has already attracted the attention of the scientific community working at the interface between soft and active matter ~\cite{BookGentile,Shankar2019HydrodynamicsOA}.

\subsection{External driving}

Before concluding, we briefly comment on the possibility of controlling defect dynamics through external driving, such as electromagnetic fields~\cite{mullol2020}.

It has been experimentally shown, for example, that 
a microtubule-based active nematic layer in contact with a field-responsive smectic liquid crystal can be aligned by a magnetic field, turning a chaotic flow into laminar \cite{Guillamat2016}. Specifically, the layered structure of the smectic phase 
allows a liquid to flow more easily along the planes than the perpendicular direction, which would respond as a solid to stresses. Thus, a magnetic field applied perpendicularly to the planes of the smectic phase would align the active nematic parallel to the planes, because of the much reduced viscosity along this direction.

In active micro-emulsions, a major challenge is represented by the poor controllability of the direction of motion, considerably hindering the use as functional materials
in several technological
applications, such as cargoes or drug delivery.
To partially overcome this limitation, external fields can be used to stabilize and control motion. In Ref. \cite{maass_arx}, for example, a magnetic field has been used to control the swimming direction of Marangoni self-propelled droplet hosting a passive nematic with an integer topological defect of charge $+1$, resulting from the perpendicular anchoring of the nematic at the droplet interface. 
Here, the droplet propulsion is due to the asymmetry of concentration gradient of a surfactant accumulated at the interface, while the direction of motion is set by a balance between magnetic field and elastic deformations of the nematic, leading to a wide range of motility regimes, including straight and oscillatory dynamics. The defect, initially at rest at the center of a quiescent droplet, moves towards the droplet anterior for a straight motion or is displaced upward for a droplet moving downward. A more complex scenario is that discussed in Ref.\cite{lancia2019}, where Marangoni droplets hosting a photo-responsive passive cholesteric liquid crystal exhibit a  conversion between right-handed and left-handed trajectories in response to light. 
Under these conditions,  the disclination lines, initially a double spiral structure at the interface due to the perpendicular anchoring of the director field, progressively unwind, thus vanishing the twisting power and producing an achiral nematic, and then reform with inverted handedness, yielding an helical trajectory with opposite chirality. This mechanisms is controlled by the photoisomerization 
of the chiral dopant which inverts its helicity under light causing the cholesteric helix to unwind through a compensated nematic state and rewind with opposite handedness.

Finally,  a particularly demanding form of control is defect braiding, in which defects are steered around one another along prescribed trajectories, rather than merely aligned or guided on average. This has recently been achieved in passive liquid crystals, in simulations~\cite{braidingtoth2026} and experiments~\cite{Leibraiding2026}. Experimentally, disclination lines entangling colloidal particles in a nematic have been woven by light into chiral double-helix structures, whose topological states can be encoded as ``nematic bits''~\cite{Leibraiding2026}. Extending this degree of control to active systems, where defects are self-propelled and continuously created and annihilated, remains an open challenge.

These experiments suggest that, although at an early stage, the use of external driving represents a promising strategy to control defect dynamics and (active) liquid crystal droplet trajectories. On the theoretical side, a computational model fully resolving order, defect, hydrodynamics and external driving is still missing, albeit the active liquid crystal theory combined with phase field models (discussed in Sec.\ref{sec:III}) and lattice Boltzmann simulations would be a natural candidate to fill this gap, as they can be straightforwardly extended to include external fields.

\section{Conclusions and perspective} 
\label{sec:VI}
In this paper we have reviewed recent advances in the field soft biological and bio-inspired materials. This fascinating kind of matter is characterized by the unique ability of consuming internal energy to perform work on the surrounding environment and, more generically, accomplish to the tasks which are necessary to preserve the life of the single constituents or a whole colony.
For this reason, biological and bio-inspired materials are often addressed as \emph{active systems}, unlike their \emph{passive} counterpart which is governed by classical mechanics and thermodynamics.

In Section~\ref{sec:II} we have reviewed some topical examples of active systems of biological origin (bacterial colonies and epithelial tissues) and artificial realizations. In this latter class we distinguished between systems whose constituents are naturally found in nature but reassembled in the lab to obtain new materials with peculiar dynamical properties, as in the case of suspensions of cytoskeletal filaments, and systems  whose elementary constituents are synthesized and engineered to mimic the behavior of biological systems, as in the case of patchy colloids and Janus particles.

Many active systems exhibit striking collective properties and often develop ordered phases where the individual constituents locally align giving rise to what are customarily called in the literature as \emph{active liquid crystal}. In Section~\ref{sec:III} we presented a mathematical approach to capture order in active systems focusing on the case of polars, nematics, and hexatics which are the most common symmetries observed in such cases. Building on the emergent orientational order, hydrodynamic theories have been developed to take into account the exchange of momentum among the active particles and the surrounding environment, as well other features like elasticity and viscosity which are standard concepts borrowed from the Landau-de Gennes nematohydrodynamic theory for passive liquid crystals.
As their passive counterpart, active liquid crystals can develop topological defects (or disclinations). These are regions where the local order is disrupted and no preferential direction of alignment can be defined. Recent research have clarified how defects play an important role in a number of biological processes, from sporulation during bacterial biofilm formation, to cell apopthosis in epithelial tissues and morphogenesis in the \emph{Hydra} development. These topics were presented in Section~\ref{sec:IV}, where we also reviewed how the active liquid crystal theory provides a qualitative description of the dynamics of active defects, and how their unbinding can lead to the stabilization of a non-equilibrium chaotic regime, namely \emph{active turbulence}. 
Finally, in Section~\ref{sec:V} we focused on three possible strategies to tame such a chaotic behaviour observed in a multitude of different realizations. The first one is to confine the active system in a geometry whose extension is comparable with the typical size spanned by the active flow. The second strategy consists instead of exploiting the topological constraints arising in peculiar geometries (like a sphere, a torus, or a disk) to stabilize topological defects even at low levels of active forcing and the third one uses external stimuli (such as light or magnetic fields) to control the dynamics of topological defects and active droplets.

The directions of future research in the field of soft biological materials and, more broadly in active matter, are rather vast and have been extensively discussed in several  reviews \cite{marchetti2013,Gompper2020,mullol2020,doostmohammadi2018}. Here, we shortly focus on some perspectives offered by the active liquid crystal theory, in combination with other continuum approaches, in tackling open problems and in identifying promising directions of research.

As discussed in this review, the active gel theory combined with phase field models and lattice Boltzmann simulations represent, so far, the most successful computational machinery to describe the mesoscale physics of active gels (even under confinement) fully incorporating ordering properties and elasticity of liquid crystal, interfacial dynamics of droplets and hydrodynamic interactions. Although the success has been demonstrated in two dimensional systems, only recently efforts have been dedicated to investigating the physics in three dimensions for selected applications \cite{majorana2024}. Indeed, while a number of studies have been focused on active polar \cite{aronson2012,tjhung,tjhung2015,ziebert2016}, and nematic \cite{ruske2021,nejad2023} to model the ordering properties of active gels in three dimensional droplets, much less is know for cholesterics \cite{tjhung2017,carenza2019}, where the interplay between the activity (extensile or contractile) and the intricate structure of disclination lines and loops is largely unexplored. In addition, a particular case of interest is that in which the helical order of cholesteric liquid crystal is in more than one direction, as typically found in blue phases liquid crystals \cite{mermin_blue,yoshizawa}. This peculiar ordering leads to the formation of a fully periodic three-dimensional network of topological defects including lines, loops and skyrmions which critically shape the mechanical properties. In the presence of activity, these systems may provide a mechanistic model of molecular motors operating on double twist DNA bundles, whose experimental evidence has been reported, for example, in Ref.\cite{lefor1994}.
In this respect a linear stability, even for two-dimensional systems, is still missing \cite{metselaar2019}, as well as an analysis of the effects of contractile and extensile stresses in three dimensions.
Fascinating avenues 
are also offered by three dimensional active multiple emulsions \cite{negro2025}, where an active gel is confined in a layer of a vesicle with topological constraints imposed by the anchoring at the fluid interfaces, as well as unexpected links between active nematics under confinement and Majorana quasi-particles as analogs of charged disclination lines \cite{majorana2024,johnson2025}, of potential relevance to topological quantum computers

Even in two dimensions, there are a number of interesting directions one may envisage. Although in this review we have mainly focused on systems displaying different types of orientational order (polar, nematic, cholesteric, hexatic) separately, in many real materials these ordering features may coexist leading to a plethora of new complex phases, such as ferroeletric and hexanematic liquid crystals \cite{chen2020,huber2018,lavren_2020,Eckert2023,armengol2023}. 
Despite the large number of examples observed in nature, only recently the active gel theory has been extended to study specific instances of nemato-polar systems
\cite{amiri2022,venkatesh2025,vafa2025,aprile2026}, by either considering two separate order parameters or by modeling 
polar and nematic interactions from competing terms in the free energy. 
The concurrent presence of integer and semi-integer defects, sometimes connected by a string, can lead to completely novel dynamic scenarios, such as  depolarization lines connecting half-integer defects and closed depolarization loops, whose study is at its infancy. Understanding, for example,
how hydrodynamic interactions impact these objects as well as how defects modify the active turbulent regime is largely unknown. In this respect, developing minimal models that incorporate different symmetries is essential for understanding the relevance  in biological systems, such as cytoskeletal networks and tissues.

Finally, a natural extension of the active liquid crystal theory is to the physics of many interacting active droplets. Although several attempts have been pursued along this direction, these are essentially confined to two dimensional periodic geometries and in the absence of hydrodynamics to model, for example, epithelia \cite{armengol2023}. However, in many relevant biological phenomena, such as tumor spreading and cell migration through interstices, confinement and flow pose additional challenges and constraints, since they profoundly impact morphology and migration mode \cite{tiribocchi2023,davidson2015,bogdan2022,au2016}.

Often, such systems consist of close-packed deformable grains whose internal length scale leads to complex rheological behaviors, including topological transitions, memory effects and avalanches. Building a computational model fully capturing the multiscale physics of such systems is currently an open problem that, besides the biological interest, may hold relevance for technological applications, from porous materials and to tissue dynamics.   

\acknowledgements
A.T. and L.N.C. acknowledge the support of CNR (Italy) and TÜBİTAK (Türkiye) through the 
project MetaCell, within the bilateral agreement for scientific cooperation.

\section*{Competing interests}
The authors declare no competing interests

\section*{Author contributions}
L.N.C. and G.N. conceived the idea and wrote the original draft. All authors discussed the results and wrote/reviewed the manuscript.

\bibliography{biblio.bib}

\end{document}